\documentclass[reprint,amsmath,amssymb,aps]{revtex4-2}
\usepackage{graphicx}
\usepackage{dcolumn}
\usepackage{bm}
\usepackage{float}
\usepackage{color}
\usepackage[dvipsnames]{xcolor}
\usepackage{hyperref}
\usepackage{graphicx}
\usepackage{braket}
\usepackage{appendix}
\usepackage{chemmacros}
\usepackage{natbib}
\usepackage{multirow}
\usepackage{amsmath}
\hypersetup{colorlinks=true, citecolor=blue, urlcolor=blue, linkcolor=blue}

\begin{document}

\preprint{APS/123-QED}

\makeatletter
\newcommand*{\balancecolsandclearpage}{%
  \close@column@grid
  \cleardoublepage
  \twocolumngrid
}
\makeatother


\title{
Tuning the band topology and  topological Hall effect in  skyrmion crystals via the  spin-orbit coupling
}   

\author{Arijit Mandal}{}\affiliation{Condensed Matter Theory and Computational Lab, Department of Physics, IIT Madras, Chennai-600036, India}
\affiliation{Center for Atomistic Modelling and Materials Design, IIT Madras, Chennai-600036, India}
\author{S. Satpathy}\altaffiliation[satpathys@missouri.edu]{}\affiliation{Condensed Matter Theory and Computational Lab, Department of Physics, IIT Madras, Chennai-600036, India}
\affiliation{Center for Atomistic Modelling and Materials Design, IIT Madras, Chennai-600036, India}
\affiliation{Department of Physics \& Astronomy, University of Missouri, Columbia, MO 65211, USA}    
\author{B. R. K. Nanda}\altaffiliation[nandab@iitm.ac.in]{}\affiliation{Condensed Matter Theory and Computational Lab, Department of Physics, IIT Madras, Chennai-600036, India}
\affiliation{Center for Atomistic Modelling and Materials Design, IIT Madras, Chennai-600036, India}


\begin{abstract}

The topological Hall effect is the result of spin-asymmetric deflection of charge carriers flowing through a non-collinear spin system.  Effective manipulation of the topological Hall conductivity (THC)  in skyrmions is currently a vigorous area of research with an eye towards potential spintronics application.
Here, we show that the band topology and the THC in a skyrmion crystal can be tuned by changing the strength of the Rashba spin-orbit coupling (SOC), which can be accomplished via a perpendicular electric field. This results in the change of the subband Chern numbers and a transition between ordinary insulator and Chern insulator as the Rashba SOC is varied. For partially filled subbands, the Rashba SOC can tune the THC and reverse its sign, 
so that the direction of the Hall current is flipped. The critical Rashba strength for this depends on the skyrmion type and the carrier density. 
We extend our analysis to the cases of Dresselhaus and Weyl SOC as well, and show that they can be directly mapped to the Rashba SOC case and therefore lead to similar results.
Our work opens up the scope to go beyond the existing avenues for the control of charge transport in skyrmion crystals.
\end{abstract}
\maketitle

\section{Introduction}
Skyrmions are topologically protected structures that have garnered significant attention in recent years for the fundamental science as well as for potential application in spintronics devices\cite{skyrme62, hlm1, bogdanov1, Roadmap, Fert, nagaosa-review, tokura-review, Beyond, Student-review}. Originally proposed by Skyrme\cite{skyrme62} as a model for baryons in particle physics in the 1960s, they have found widespread application in condensed matter physics including Bose-Einstein condensates\cite{BEC1,BEC2}, liquid crystals\cite{liquid-crystal}, quantum Hall magnets\cite{qhe1,qhe2}, and helimagnets\cite{hlm1, hlm2}. The magnetic skyrmions that form in the solid are nanometer-size twirling spin structures, which are metastable due to the topological protection
with a fixed quantized winding number.

Since the prediction of the stable skyrmion state in magnetic solids\cite{hlm1}, skyrmions have been observed in a plethora of materials \cite{tokura-review} such as the ferromagnetic semiconductors with B20 structure such as MnSi and FeGe\cite{skx2, semicon-skx}, transition metal oxides \cite{the-com2, the-com3}, Heusler alloys \cite{heusler}, Janus materials \cite{janus-skx}, moir\'e heterostructures \cite{moire-skx}, intermetallic compounds \cite{the-com8}, and magnetic multilayers \cite{skx5}.
Skyrmion crystals, the ordered arrays of skyrmions, have been suggested theoretically\cite{skx-dmoria} in magnets as well as in two-dimensional electron gas\cite{Brey}. Several theoretical studies have also examined the energetics of the formation of stable skyrmion crystals via Dzyaloshinskii–Moriya interaction, magnetic frustration, anisotropy, etc.\cite{hlm2, skx-dmoria, multiple-q, skx-frus, skx-frus-1, anisotropy1, anisotropy2, hayami-four-spin, Paul-four-spin, skx-theory}

Experimentally, skyrmion crystals have been observed in several systems using neutron scattering and Lorentz transmission electron microscopy. 
A hexagonal skyrmion lattice was first observed in  MnSi \cite{skx2} and later in other intermetallics such as  FeCoSi, FeGe, and Co$_8$Zn$_{10}$Mn$_2$ \cite{skx7, skx3, skx4, skx6}.
Other types of lattices have been found as well, viz., the triangular lattice in  Gd$_2$PdSi$_3$, Gd$_3$Ru$_4$Al$_{12}$, and the CoZnMn alloy system\cite{skx8, the-com8, skx12},
as well as the square lattice in GdRu$_2$Si$_2$, EuAl$_4$, and
in monolayer Fe/Ir($111$) \cite{skx5, skx9, skx10}.

The transport of electrons in the presence of skyrmions has emerged as an important area of research.
Included in this are the topological Hall effect (THE) and the spin Hall effect (SHE) as well as the skyrmion Hall effect, which concerns the transport of the skyrmions themselves\cite{thc-review}.
THE arises from the transverse motion of electrons due to the emergent magnetic field of the noncollinear spin texture, an effect akin to the anomalous Hall effect (AHE) in the collinear spin system.

\begin{figure}[hbt!]
    \centering
    \includegraphics[width=1\linewidth]{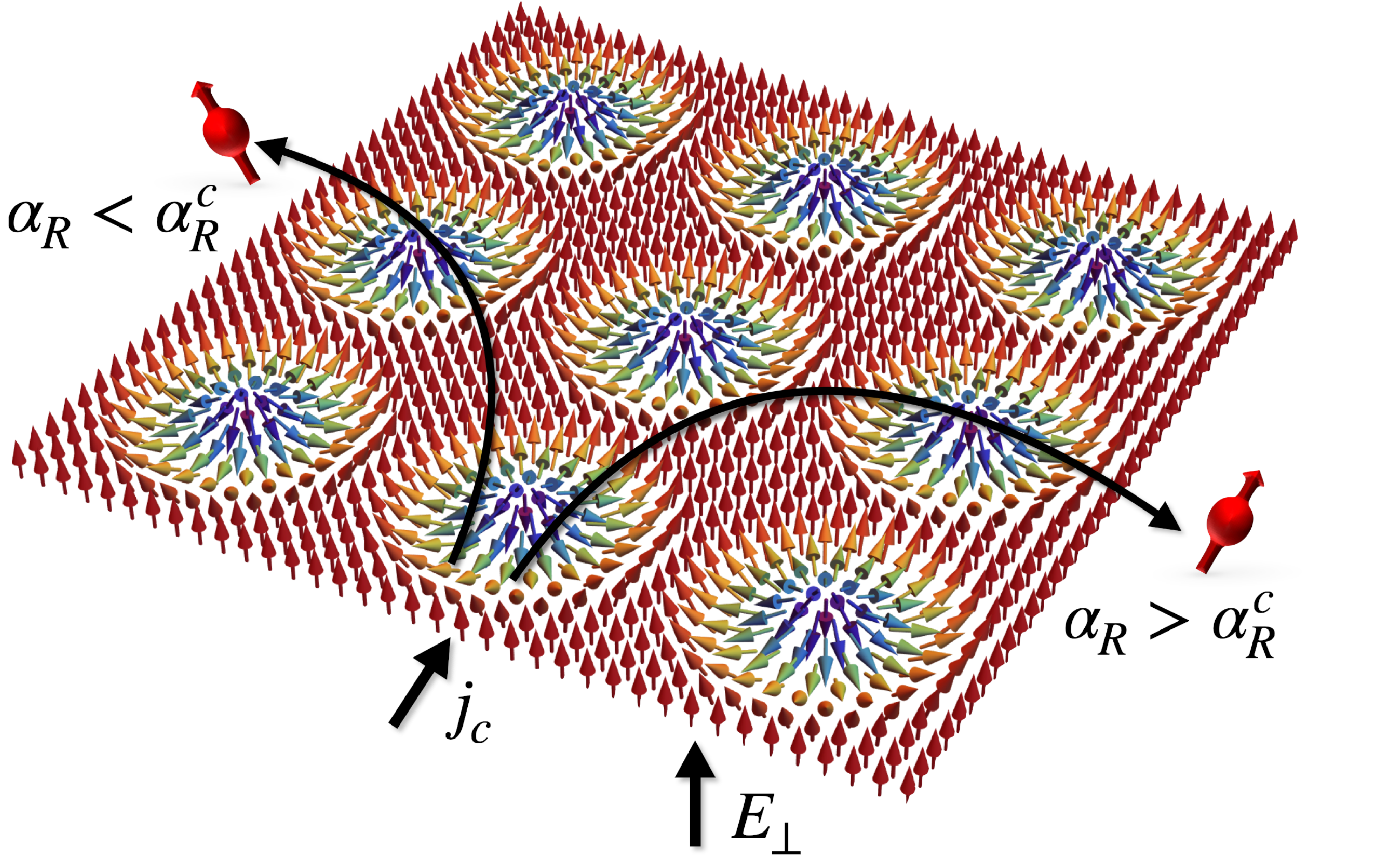}
    \caption{Topological Hall effect in the skyrmion crystal. The magnitude of the Hall current can be changed and its sign can be reversed by tuning the Rashba SOC through a critical value $\alpha_R^c$. The Rashba strength $\alpha_R$  can be modified via a perpendicular electric field $E_\perp$ or by growing an overlayer. Similar effects can be obtained with the Dresselhaus and Weyl SOC also as discussed in the text.
}
    \label{im1}
\end{figure}

\begin{figure}[hbt!]
    \centering
   \includegraphics[width=0.6\linewidth]{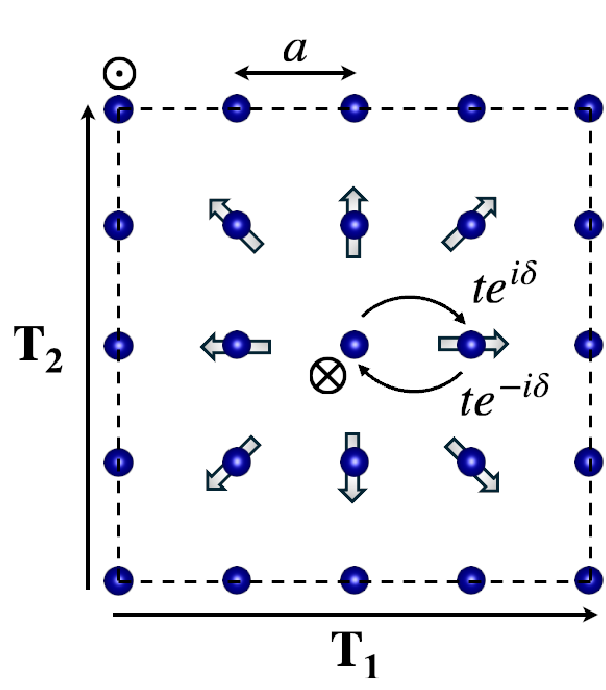}
    \caption{
    The $4 \times 4$ unit cell  used in our calculation. Each site is a magnetic atom with its spin oriented according to the spin texture Eq. \ref{spin texture}, with the origin being the center of the square. Together, the sixteen sites describe a single skyrmion, which is repeated via the lattice translation vectors $\bold {T_1} $ and $ \bold {T_2}$ forming thereby the skyrmion crystal. In the figure, the central spin points into the page, the boundary spins point out of the page, while the rest of the spin directions are shown only schematically. The electron hops from atom to atom with a complex hopping amplitude due to the emergent magnetic field. 
    }
    \label{unit-cell}
\end{figure}

\begin{figure*}[hbt!]
    \centering
   \includegraphics[width=0.65\linewidth]{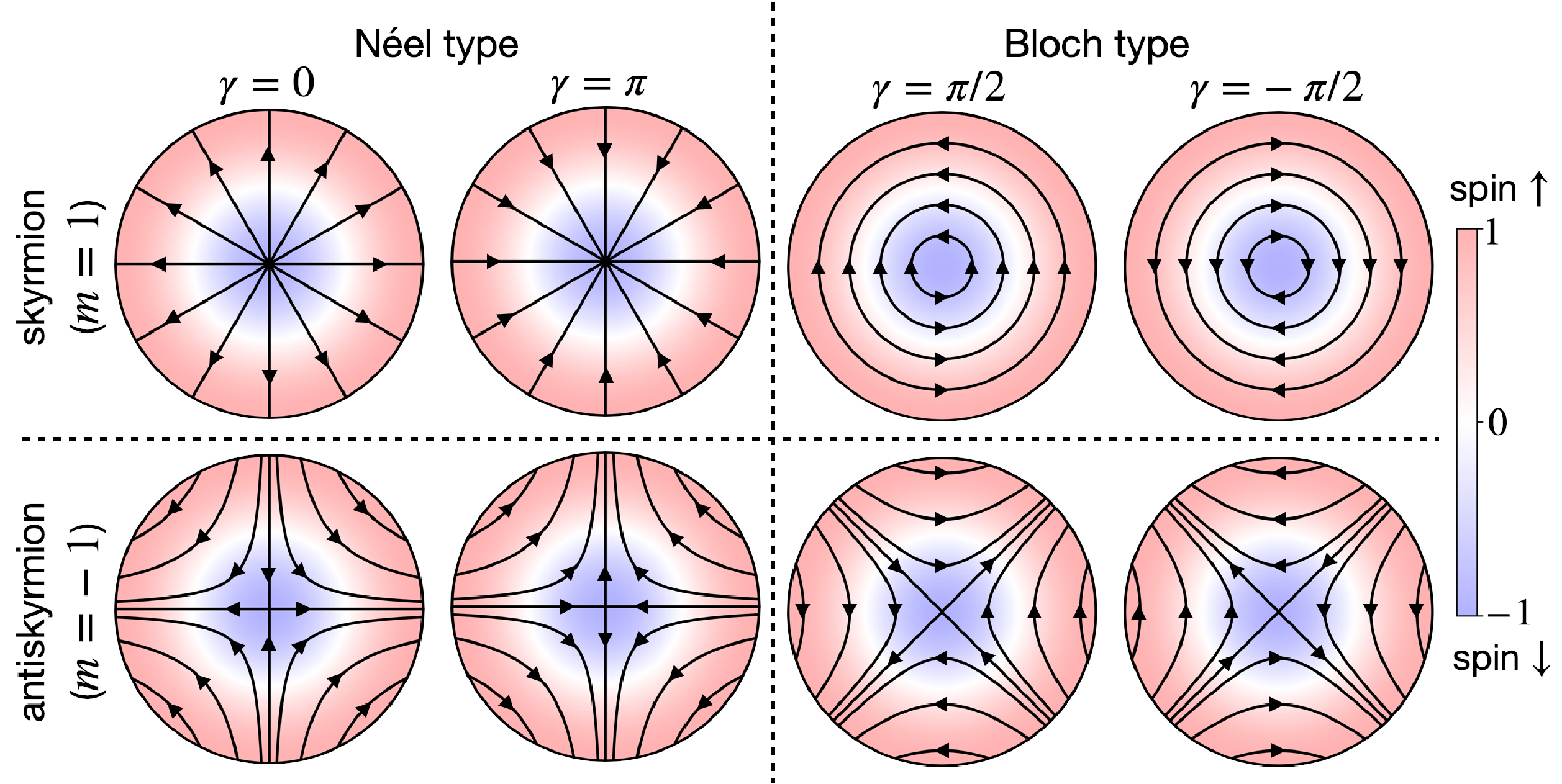}
    \caption{Illustration of the skyrmion spin texture using field lines. The tangent indicates the planar component of the local skyrmion spin, while the color
    indicates the component normal to the plane.
    The spin at the origin is pointed down, changing gradually to the spin up orientation at the skyrmion boundary, as indicated by the color coding.
    }
    \label{skyrmion}
\end{figure*}


There has been mounting interest in the electrical field control of the Hall effects because it would enable spin manipulation without an external magnetic ﬁeld or a large current injection.  However, there has been limited success in this direction using conventional field-effect structures involving magnetic materials adjacent to gate dielectrics. More recently, it has been suggested that this may be achieved through the spin-orbit coupling (SOC) effect, which may be modified by an applied electric field. In a recent experiment, it was found that both the AHE and the THE (arising from the interfacial skyrmions) can be 
tuned by an applied electric field in the oxide heterostructure via the SOC effect of an inserted layer\cite{oxide1, oxide2}. 

In this work, we study the THE in the skyrmion crystal in the presence of three types of SOC, viz., the Rashba, Dresselhaus, or the Weyl type. The Rashba case is studied in some detail, while we show that the other two cases can be mapped to the Rashba case for the skyrmion types we have considered here. Quite remarkably, we find that not only can the magnitude of the topological Hall conductivity (THC)  be significantly changed by varying the strength of the Rashba SOC via an applied electric field, but also that the sign of the Hall conductivity can be reversed by varying the Rashba SOC around a critical value. The band topology in the skyrmion crystal can also be changed by varying the strength of the SOC term, resulting in a change in the subband Chern numbers, which can lead to a flipping of the direction of the topological edge current.

\section{Model Hamiltonian and the electron band structure}
We consider the model Hamiltonian for a square lattice for an electron moving through the skyrmion crystal in the presence of the Rashba SOC 
\begin{eqnarray}
    \mathcal{H} &=& 
     - t \sum_{\langle ij \rangle}  (c_{i}^{\dagger} c_{j} + c_{j}^{\dagger} c_{i})
     -J\sum_{i} \boldsymbol{S}_i \cdot c_{i}^{\dagger}\boldsymbol{\sigma}c_{i} \nonumber \\
     &-& i \alpha_R \sum_{\langle ij \rangle} 
     c_{i}^{\dagger} (\boldsymbol{\sigma} \times \boldsymbol{r}_{ij})_z c_{j} + H. c..
    \label{eq1}
\end{eqnarray}
The first term is the nearest neighbor hopping term, where 
we have suppressed the spin index, with 
$c_i^\dagger \equiv (c_{i\uparrow}^\dagger,  c_{i\downarrow}^\dagger)$ and 
$c_i \equiv (
c_{i\uparrow} , c_{i\downarrow})^T$, $i$ is the site index on the lattice, $\sigma$ is the spin index,  
$c_{i\sigma}^{\dagger}/ c_{i\sigma}$ are the corresponding 
electron creation/annihilation operators,   $\langle ij \rangle$ indicates distinct pairs of nearest neighbors, and H. c. indicates  Hermitian conjugate of the last term only.
Note that the sites represent individual magnetic atoms in the structure and several atoms taken together (16 here) form a single skyrmion, which is periodically repeated to make the skyrmion crystal (see Figure \ref{unit-cell}).
The second term is the exchange interaction between the site-fixed spins $\boldsymbol{S}_i$, commonly taken as classical spins, 
with the skyrmion texture and  $\boldsymbol{\sigma}$ is the Pauli matrix describing the electron spin. 
The third term is the lattice version of the Rashba SOC\cite{Rashba_ref, Kane-Mele}, which breaks the $ z \rightarrow -z$ symmetry. 
In this term, $\alpha_R$ is the Rashba interaction strength and $\bold r_{ij} = \bold{r}_j-\bold{r}_i$ is the vector distance between the nearest-neighbor sites. 
It is easily shown that 
for the free-electron case (small $k$ limit of the lattice case), it yields the well-known form of the 
Rashba Hamiltonian ${\cal H}_R^k = 
\alpha_R' (\boldsymbol{\sigma} \times \boldsymbol{k}) \cdot \hat z$,
with a scaled coefficient $ \alpha_R'= 2a^2 \alpha_R  $, where $a$ is the lattice constant \cite{supplementary}.
The Rashba SOC can arise due to the symmetry breaking by an overlayer or by applying an external electric field.

 For the free electrons in the presence of an electric field, the strength of the Rashba term is given by 
$\alpha_R' = \hbar^2E_z/2m_e^2c^2$. This expression however severely underestimates the magnitude of the Rashba SOC, if we take $E_z$ to be the surface electric field. The reason for this discrepancy\cite{Shanavas} is that the relativistic effects originate largely from the nuclear region, and the role of the surface electric field is to modify the electron density near the nucleus by mixing the orbitals due to broken symmetry. This is a much larger effect than simply computing a free-electron-like expression due to the surface electric field. 

The exchange interaction is often taken as $J \rightarrow \infty$, which is reasonable considering that for typical solids $ J \sim 2-3 $ eV, while the typical hopping is a fraction of an eV.  Also, the essential physics discussed here is not changed by taking this limit. However, there are no additional complications if a general  $J$ is used. As an example, the band structure and THC for finite $J$ are shown in the Supplementary Material \cite{supplementary}.  
In infinite $J$ limit, the problem reduces to that of the spinless fermions, since the electron state with spin anti-aligned with the lattice spin is forbidden. The spinless fermion operators are described by
 \begin{equation}
    d_i^\dagger = \cos (\theta_i/2) \ c_{i\uparrow}^\dagger +
    e^{i\phi_i}\sin(\theta_i/2) \ c_{i\downarrow}^\dagger,
\end{equation}
 where $(\theta_i, \phi_i)$ is the orientation of the lattice spins
in spherical coordinates.
The Hamiltonian assumes the form
\begin{eqnarray}
    \mathcal{H}_R &=& -t\sum_{\langle ij \rangle} \cos{\frac{\theta_{ij}}{2}}e^{i a_{ij}} \ d_i^\dagger d_j \nonumber \\
    &-& i \alpha_R \sum_{\langle ij \rangle} \big[ e^{i\phi_j}(r_{ij}^y + i r_{ij}^x) \cos{\frac{\theta_i}{2}}\sin{\frac{\theta_j}{2}} \nonumber \\
    &+& e^{-i\phi_i}(r_{ij}^y - i r_{ij}^x) \cos{\frac{\theta_j}{2}}\sin{\frac{\theta_i}{2}} \big] \ d_i^\dagger d_j + H. c.,
    \label{local-inf-model}
\end{eqnarray}
where the electronic hopping is reduced by the factor $\cos (\theta_{ij}/2)$ and the Peierls phase factor $e^{i a_{ij}}$
indicates the presence of the emergent magnetic field, which turns the spin of the electron as it moves from site to site.
In Eq. (\ref{local-inf-model}), 
$\theta_{ij} = \cos^{-1} ( \sin{\theta_i}\sin{\theta_j} \cos{(\phi_i - \phi_j)} + \cos{\theta_i}\cos{\theta_j} ) $
and 
$a_{ij} = - \tan^{-1} \big[ \sin(\phi_i - \phi_j) / 
(\cot(\theta_i/2) \cot(\theta_j/2) + \cos(\phi_i-\phi_j) )\big] $,
and the subscript in  $\mathcal{H}_R$ has been added to distinguish it from the Dresselhaus and Weyl Hamiltonians, $\mathcal{H}_D$ and $\mathcal{H}_W$, discussed later.
In an earlier work\cite{Sahu}, we have found that the presence of the Rashba or the Dresselhaus SOC affects the electron motion in such a way that the electron can form a self-trapped bound state at the skyrmions core, if the conditions are right. 

We take the spin texture of the isolated skyrmion in the form 
\begin{eqnarray}
    \theta(r) &=& 
    \begin{cases}
    \pi(1-r/\lambda) & \text{for}~ r \leq \lambda \\
    0 & \text{for}~ r > \lambda,
    \end{cases}\nonumber \\
    \phi(\alpha) &=& m\alpha + \gamma,
    \label{spin texture}
\end{eqnarray}
where $(\theta, \phi)$ are the polar and azimuthal angles in the spherical polar coordinates and $(r, \alpha)$ is the 2D position of the lattice spin in polar coordinates with the origin coinciding with the skyrmion centre.
Here, $\lambda$ is the skyrmion radius, $m$ is the winding number or the vorticity, and $\gamma$ is the helicity of the skyrmion. For a single-turned skyrmion and antiskyrmion, $m$ takes the value of $+1$ and $-1$, respectively. Depending on the helicity,  skyrmions can be classified as N\'eel  ($\gamma=0\ {\rm or}\  \pi$) and Bloch types ($\gamma=\pm \pi/2$) as illustrated in Figure \ref{skyrmion}. 

We consider a square lattice of localized spins with a magnetic unit cell of $N \times N$ sites hosting a skyrmion 
with the spin texture Eq. \ref{spin texture}.
Thus the skyrmions form a square lattice with 
lattice constant $Na$, where $a = 1$ is the distance between the neighboring spins.
We took $N = 4$ with the skyrmion radius $\lambda = 2a$ in our calculation, so that there are $N^2 = 16$ sites in the magnetic unit cell of the skyrmion crystal. 
The unit cell of the skyrmion crystal is shown in Figure \ref{unit-cell}.
Here, we have used a smaller ($4 \times 4$) magnetic cell for illustration, but the physics remains unchanged if a larger cell is used, since the total flux through the cell remains the same. Some results for the 
$10 \times 10$ are shown in the Supplementary Materials \cite{supplementary}.

The SOC modifies the emergent magnetic field of the skyrmion, which we discuss below. In addition, it modifies the magnitude of the effective hopping \cite{supplementary}. Both these effects modify the electronic band structure as well as the topological properties.

{\it Emergent magnetic field}.
Here, we compute the emergent magnetic field seen by the electron for an isolated skyrmion in the presence of the Rashba SOC. The results will be needed in the interpretation of the THC,
particularly the quantum Hall plateaus, as the SOC strength is varied.

In the lattice models with a magnetic field, 
hopping integrals between lattice sites acquire a complex phase factor 
$t \times \exp \big[\frac{-ie}{\hbar} \int_{a}^{b}\bold{A}.d\bold{r}\big]$ via the Peierls substitution, where $t$ is the hopping without the magnetic field from site $a$ to $b$, $\bold{A}$ is the vector potential, 
$-e <0$ is the charge of the electron, and 
the integral is taken over the path joining the two sites. 
To compute the magnetic field, we take an infinitesimal square loop around the point $\bold r$ and compute the flux through it using the Peierls phase factors accumulated by the electron as it travels around the square loop, following the Hamiltonian Eq. (\ref{local-inf-model}). 
After straightforward but tedious algebra \cite{supplementary}, we find that  the magnetic field, which is normal to the skyrmion plane, is given by
\begin{widetext}
\begin{eqnarray}
    B_{SK}^R(r) = 
    \begin{cases}
    \frac{\Phi_0}{2\pi} \times  \big[\frac{ \pi}{2\lambda r}\sin\frac{\pi r}{\lambda} 
     - \frac{\alpha_R }{t} ( \frac{\pi}{\lambda}\cos{\frac{\pi r}{\lambda}} + \frac{1}{r} \sin\frac{\pi r}{\lambda} ) \cos \gamma \big] & \text{for}~ r < \lambda \\
    0 & \text{for}~ r > \lambda,
    \end{cases}
    \label{sk-rb-mag}
\end{eqnarray}

where $\Phi_0 = h/e$ is the  flux quantum. 
The superscript in $B_{SK}^R$ refers to the Rashba SOC case.
The first term is due to the skyrmion alone, and it is independent of the helicity $\gamma$, but depends on the vorticity ($m = +1$ here), while the second term is the modification of the magnetic field due to the Rashba SOC. The $B_{SK}^R$ has a discontinuity at the boundary which arises due to the discontinuity in the derivative of the spin texture Eq. \ref{spin texture}. This can be understood from the magnetic field expression of Eq. \ref{B2} discussed later.\\

For the antiskyrmion, the magnetic field due to Rashba SOC is
\begin{eqnarray}
    B_{AS}^R(r, \alpha) =
    \begin{cases}
    \frac{\Phi_0}{2\pi} \times  \big[- \frac{ \pi}{2\lambda r}\sin\frac{\pi r}{\lambda} - \frac{\alpha_R }{t} ( \frac{\pi}{\lambda}
   \cos{\frac{\pi r}{\lambda}} - \frac{1}{r} \sin\frac{\pi r}{\lambda} )  \cos({2\alpha} - \gamma) \big]& \text{for}~ r < \lambda \\
    0 & \text{for}~ r > \lambda,
   \end{cases}
   \label{ask-rb-mag}
\end{eqnarray}
\end{widetext}
Note that here the magnetic field is no longer circularly symmetric. 
It can be shown by direct integration that the total
flux passing through a single syrmion/ antiskyrmion is 
$\pm \Phi_0$, irrespective of the strength of the Rashba term. 
In other words, the Rashba term does not change the total flux through the skyrmion, rather, it redistributes the flux within the skyrmion region, so that the total flux remains unchanged. The same is also true for the Weyl and Dresselhaus SOC as well.

\begin{figure} [tbh!]
    \centering
 \includegraphics[width=1.0\linewidth]{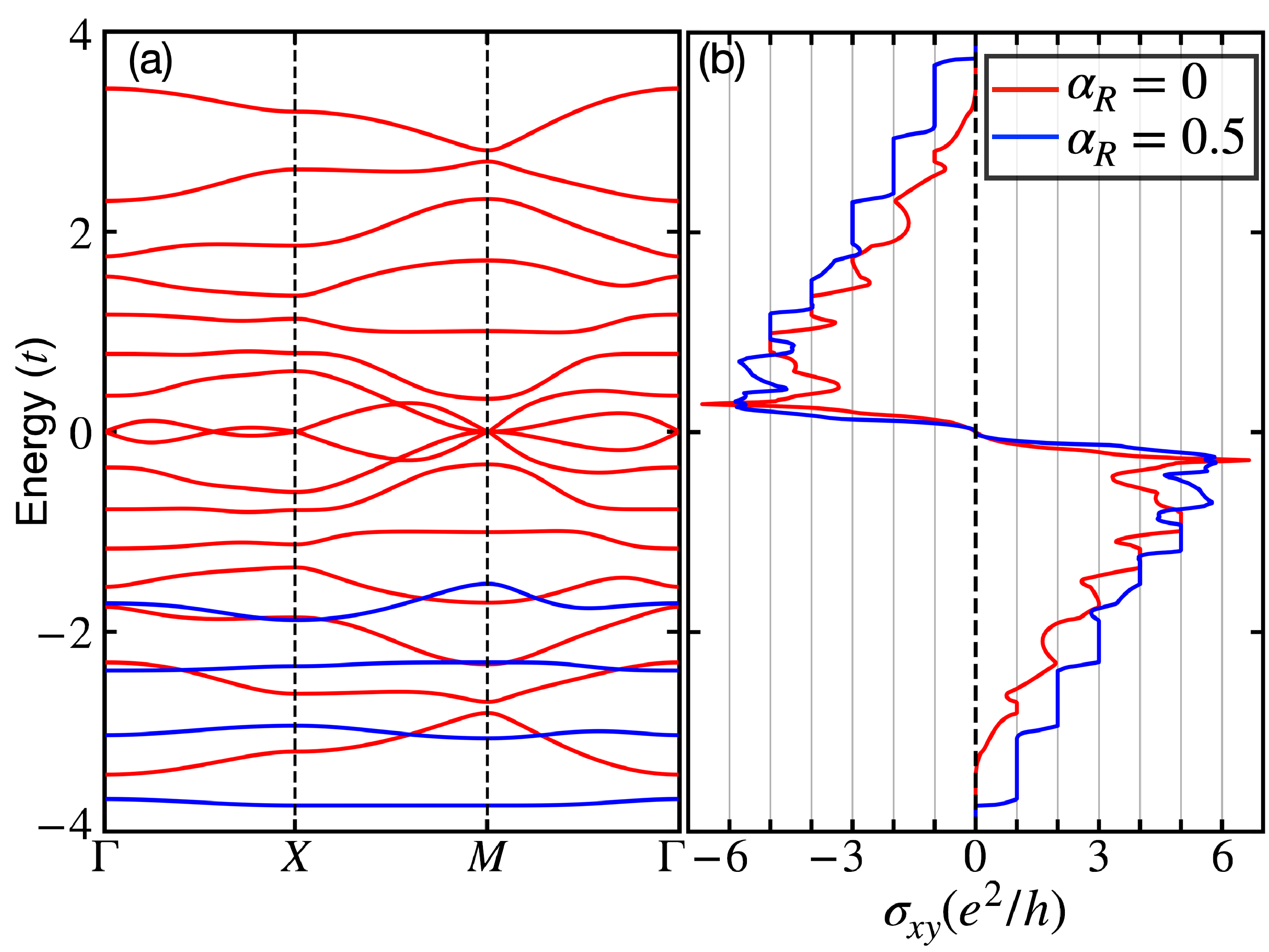}

    \caption{Band structure and THC  in the skyrmion crystal with the N\'eel skyrmion. (a) Band structure without  (red lines) and with  (blue lines) Rashba SOC. For the latter case, only the lowest four bands are shown. (b) THC as a function of the Fermi energy $\varepsilon_F$, indicating prominent Hall plateaus for $\alpha_R = 0.5 t/a$, chosen so as to produce a maximally uniform magnetic field (see Figure \ref{im2-2-1}). Here and throughout the paper, the Hund's coupling is taken to be $J \rightarrow \infty$ as discussed in the text. 
    }
    \label{im2}
\end{figure}

\begin{figure} [tbh!]
    \centering
    \includegraphics[width=1.0\linewidth]{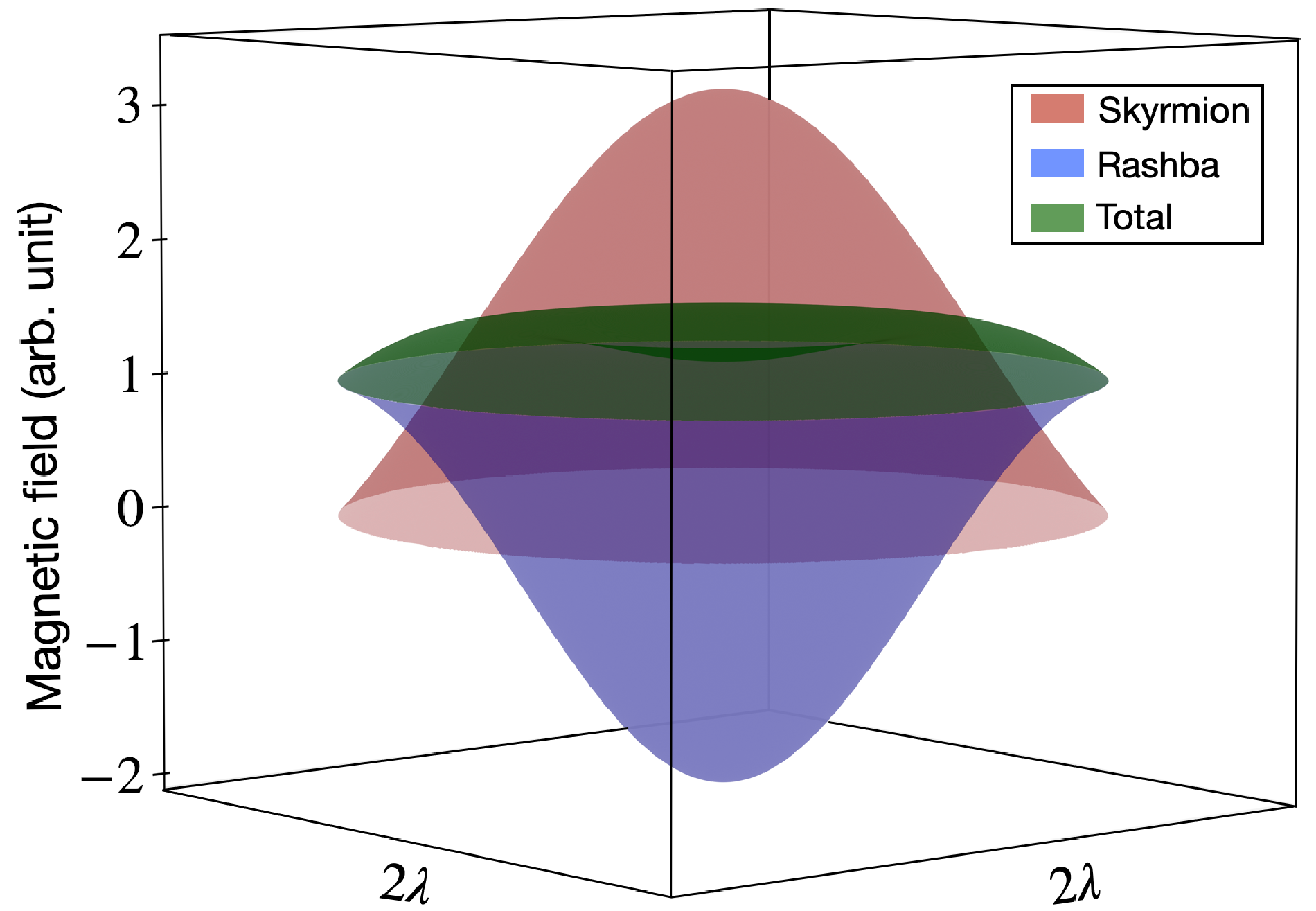}
    \caption{
    Emergent magnetic field, in arbitrary units, from Eq. (\ref{sk-rb-mag}) due to the N\'eel skyrmion (red) and the Rashba SOC term with $\alpha_R = 0.541/\lambda$ (blue), which add up to a near-uniform total magnetic field (green).
    }
    \label{im2-2-1}
\end{figure}

{\it Electron band structure}.
Electron states in a lattice with a uniform magnetic field has been long studied since the seminal work of Hofstadter\cite{Hofstadter} that produced the celebrated ``Hofstadter butterfly" graph for the allowed energy regions. In our problem, we have a non-uniform magnetic field and also we are interested in the band structure and its topological properties.
This has been examined in the literature for several lattices starting with the 
pioneering work of Hamamoto et al.\cite{nagaosa}, where the connection between the subbands and the Landau levels for the free electrons in a uniform magnetic field was made. The subband energies for the lattice case can be estimated\cite{qhe-gobel} by using the Bohr-Sommerfeld semiclassical quantization rules, which was originally applied by Onsager\cite{Onsager} to discuss the da Haas-van Alphen oscillations in solids.
Our work extends these ideas to systems where the SOC is present. 
We consider three types of SOC terms, viz., the  Rashba, Dresselhaus, and Weyl SOC, which may be engineered to be present under appropriate circumstances.
For example, it has been demonstrated that the relative strengths of the Rashba and Dresselhaus SOC terms can be tuned experimentally in semiconductor quantum wells
via interface engineering\cite{Rashba-Dresselhaus-2016} and electric field\cite{rb-elec}. 

We note that the electron states in the skyrmion crystal are in many ways similar to the electron states in the integer quantum Hall effect (IQHE), in the sense that while the electrons are subjected to a uniform magnetic field in the IQHE, for the skyrmion crystal, the emergent magnetic field is non-uniform and varies spatially. The band structures as well as the topological properties for the two cases are similar (a specific example is given in the Figure 1 of the  Supplementary Materials \cite{supplementary}).

We first consider the square lattice of single-turned N\'eel type skyrmion with helicity $\gamma=0$. The band structures with and without a Rashba SOC are shown in Figure
\ref{im2}.
The non-uniform magnetic field produces a larger subband dispersion (red curves in Figure \ref{im2} (a))  as compared to the case of the uniform magnetic field. 
The Rashba term modifies the emergent magnetic field, as discussed already, which
in turn modifies the band structure. An interesting situation arises, where by adjusting the strength of the Rashba term, one can make the emergent magnetic field
nearly uniform. In that case, the band structure will resemble the Landau level bands for the uniform magnetic field. 

To obtain this optimal value $\alpha_R^0$ that produces the maximally uniform magnetic field, we have minimized the smoothness integral 
$\int_0^{\lambda} 2\pi r \left( d B_{SK}(r)/d r \right)^2 dr$
over the skyrmion radius
using the total magnetic field Eq. (\ref{sk-rb-mag}). 
The result is:
\begin{equation} 
\alpha_R^0 = (\pi t/\lambda) (2c-1) (\pi^2-2+6c)^{-1}
\approx 0.54 \ t/\lambda, 
\label{uniform-B}
\end{equation}
where 
$c=c_1 - c_2 + \ln \ (2 \pi)$, $c_1 \approx 0.577$ is the Euler-Mascheroni constant, and $c_2 = - \int_{2\pi}^{\infty} dx \ \cos x/x  \approx -0.022 $ is the cosine integral. 
From an inspection of Eq. (\ref{sk-rb-mag}), it is obvious that the magnetic field can never be uniform for all $r$, but for the optimal Rashba strength $\alpha_R^0$, we find that the rms deviation of $B (r)$ from the mean is as small as $\approx 8 \%$.
Figure \ref{im2-2-1} shows the magnetic field for the optimal value $\alpha_R^0$. 

The optimal value expression for $\alpha_R^0$, Eq. (\ref{uniform-B}), was obtained in the continuum limit. 
However, in the skyrmion crystal, because only a small number of lattice points sample the skyrmion magnetic field (for our square lattice there are only $4 \times 4$ lattice points in the magnetic unit cell containing a single skyrmion), the optimal $\alpha_R^0$ is different from the predicted value, which is more accurate for very large magnetic cells.
Test calculations with $100 \times 100  $ unit cell yielded the $\alpha_R^0$ value as predicted by Eq. (\ref{uniform-B}).
In the present case of the $4 \times 4$ magnetic unit cell, $\alpha_R^0$ is found to be about twice of Eq. (\ref{uniform-B}). 

The  band structure for the $4 \times 4$ lattice with the Rashba strength $\alpha_R^0$ is shown in
Figure \ref{im2} (a) (in blue), which  is very similar to the bands with a uniform
magnetic field \cite{supplementary}. 
As discussed in the next Section, we show that the topology of the bands becomes modified in the presence of the Rashba SOC term, which in turn leads to the tuning and sign reversal of the THC.

\begin{figure*}
    \centering
     \includegraphics[width=1.0\linewidth]{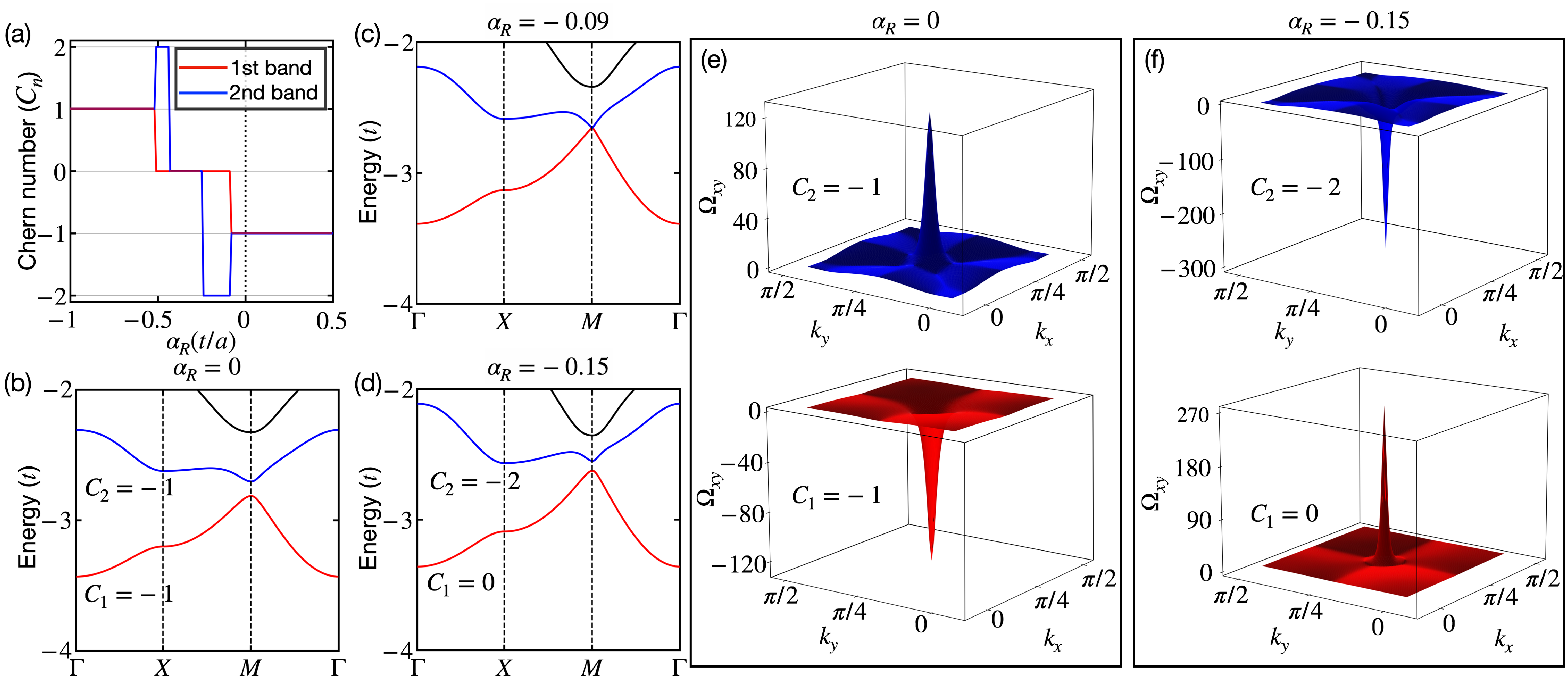}
    \caption{Tuning the band topology with the Rashba SOC for the N\'eel type skyrmion crystal. (a) The Chern number of the lowest two subbands as a function of $\alpha_R$. (b, c, d) Evolution of the band structure with $\alpha_R$ changing through the crossover point $\alpha_R \approx -0.09$. 
    The lowest two subbands touch and separate, and their Chern numbers change as indicated in (b) and (d), where we have shown the Chern numbers of the two lowest subbands. 
    (e, f) Berry curvatures for the lowest two subbands for  $\alpha_R$ just above and below the crossover point, corresponding to the band structure of (b) and (d) respectively.  
    }
    \label{chern}
\end{figure*}

\section{Topological Hall Effect with Rashba SOC}
The THC can be computed  as a sum of the Berry curvature in the momentum space following the standard expression \cite{prl-berry}
\begin{eqnarray}
    \sigma_{xy} = -\frac{e^2}{\hbar}\frac{1}{N_k A}\sum_{n\boldsymbol{k}} f(\varepsilon_{n\boldsymbol{k}})\Omega_n^z (\boldsymbol{k}),
    \label{thc}
\end{eqnarray}
where $N_k$ is the number of k-points in the BZ, $A$ is the area of the unit cell, $\varepsilon_{n\boldsymbol{k}}$ is the band energy, and $f(\varepsilon_{n\boldsymbol{k}})$ is Fermi-Dirac distribution function.
The Berry curvature $\vec \Omega (\vec k)$ enters into the semiclassical equation of motion for the Bloch electron  
and affects the motion of the wave packet if it is non-zero. 
The  velocity of the Bloch electron wave packet  is given by the expression\cite{prl-berry}
\begin{equation} \label{vel}
 \dot  {\vec  r}_{n \vec k} = \hbar^{-1}[\vec \nabla_k \varepsilon_{n \vec k} +e \vec E \times \vec \Omega_n (\vec k)], 
\end{equation}
where the second term on the right hand side is the ``anomalous velocity" $\vec V_{n \vec k}  = (e/\hbar) \vec E \times \vec \Omega_n ({\vec k})$. 
The Berry curvature $\Omega_n^z (\boldsymbol{k})$  of the $n$-th band may be evaluated from the Kubo formula
\begin{equation}
    \Omega_n^\gamma (\boldsymbol{k}) = -2 \hbar^2 \  \text{Im} \sum_{m \neq n} \frac{\langle \psi_{n\boldsymbol{k}} | v_\alpha |\psi_{m\boldsymbol{k}} \rangle \langle \psi_{m\boldsymbol{k}} | v_\beta |\psi_{n\boldsymbol{k}} \rangle}{(\varepsilon_{m\boldsymbol{k}} - \varepsilon_{n\boldsymbol{k}})^2},          
    \label{berry_curv}
\end{equation}
where the velocity operator $v_\alpha = (\hbar^{-1})\  \partial {\cal H}/\partial k_\alpha$,  $(\alpha, \beta, \gamma)$ are cyclic permutations of $(x, y, z)$,
and $\varepsilon_{m\boldsymbol{k}}$ and $\psi_{m\boldsymbol{k}}$ are the band energies and the wave functions.
In the rest of the paper, we drop the subscripts in the THC $\sigma_{xy}$ and simply call it $\sigma$.

The Chern number, defined as the sum of the Berry curvature over the Brillouin zone
\begin{equation}
\mathcal{C}_n = \frac{1}{2\pi}\int_{BZ} d \boldsymbol k\ \Omega_n^z(\boldsymbol{k}),
\end{equation}
is an important integer quantity that characterizes the topology of the band, $n$ being the band index. 
For a completely filled band, its contribution to the Hall conductivity is simply $\sigma = - (e^2 / \hbar)\  \mathcal{C}_n$, and $\mathcal{C}_n \ne 0$ indicates a non-trivial topological band.
%
\begin{figure*} [tbh!]
    \centering
     \includegraphics[width=0.8\linewidth]{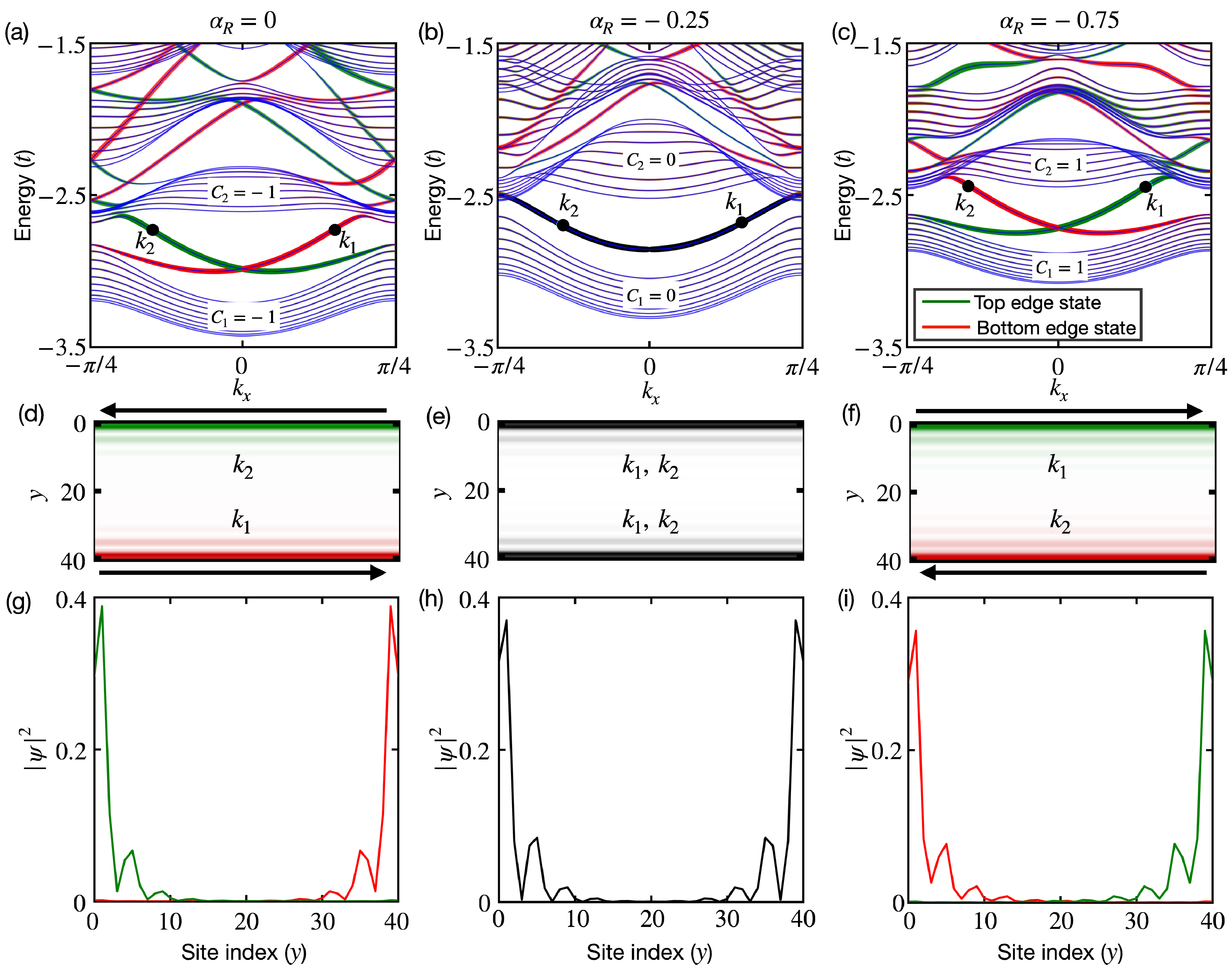}
    \caption{Edge states for the nanoribbon geometry for three different $\alpha_R$: (a) $\alpha_R = 0$, (b) $\alpha_R = -0.25$, (c) $\alpha_R = -0.75$. 
    The band topology for the three $\alpha_R$ are different, as indicated from the Chern numbers of the lowest two subbands, which vary between -1, 0, and +1  for the three cases shown.
    The ribbon width is finite along the $y$ direction (40 lattice sites corresponding to 10 magnetic unit cells), while its length is infinite along $x$. 
    The middle panel shows the  localization of the edge states either on the top edge  or the bottom edge of the ribbon,
   and the color density is proportional to  the wave function character $|\psi(k)|^2$ for the  edge states, 
   which are labelled $k_1$ and $k_2$  in the top panel. 
   The color coding, red or green, corresponds to localization along the bottom edge or the top edge of the ribbon, respectively, while
the long arrows indicate the direction of propagation of the edge states.
    The bottom panel shows the wave function characters of the same edge states,
    which exponentially decay into the bulk.  
      For  $\alpha_R = -0.25$, the two midgap states at each $k$ point are degenerate, corresponding to a state on the bottom edge and a state on the top edge, and constitute ordinary surface states with no topological character. They are shown in black colors in (b), (e), and (h).
      }
    \label{Edge}
\end{figure*}

The THC in the skyrmion crystal without the Rashba SOC has been studied by earlier authors for several lattices. The results for the square lattice are shown in the Figure \ref{im2} both without and with the Rashba SOC term. 
When the Rashba SOC is absent, the Chern number for each of the subband is $\mathcal{C}_n$ of $-1$ as in the case of LLs arising from the uniform magnetic field \cite{nagaosa, qhe-gobel}, except
for the cluster of subbands around the van Hove singularity at $E = 0$, which have the collective Chern number of $N-m$ as a group, where $N$  is the total number of subbands (16 here) and $m$ is the number of subbands in the cluster. 
If the Fermi level lies within a band gap, the THC is simply the negative of the sum of the Chern numbers of the lower bands in units of $e^2/h$. Due to the overlapping bands, these plateaus are not always visible.

With the introduction of the Rashba SOC, the band dispersion is modified, and at an optimal value $\alpha_R^0$, the subbands become nearly dispersionless as discussed earlier. At this point, the THC quantization plateaus prominent, as seen in the case of quantum Hall effect (see Figure \ref{im2}(b)). Due to the quantization in the Hall conductivity, it is sometimes referred to as quantized topological Hall effect (QTHE)\cite{nagaosa}.

Not only the band dispersion is modified due to the presence of Rashba SOC, but the topology of the band structure is also altered, leading to a change in the Chern numbers.  The Chern numbers still remain integers for the isolated individual subbands, but their magnitudes change as one varies the strength $\alpha_R$ of the Rashba SOC term.  The results are shown for the $4 \times 4$ square lattice in Figure \ref{chern} (a) for the lowest two subbands, where we see that the Chern numbers flip flop between the values $\pm 2$ for the range of $\alpha_R$ considered. 
Figure \ref{chern} (b-d) show the band structures as one passes through one of the crossover points, where the Chern number changes abruptly. The corresponding Berry curvatures are also shown in the Figure for $\alpha_R$ slightly below and above this crossover value. 
In essence, the change in the band topology occurs due to the changed magnetic field profile caused by the Rashba SOC term. 

Finally, we note that when the topological character (Chern number) of a band changes as a function of any variable ($\alpha_R$ here), the system must go through a metallic state, as seen from Figure \ref{chern} (b), (c), and (d). 

{\it Nanoribbon states}.  The varying nature of the band topology (Chern numbers) with changing $\alpha_R$ shows up in the edge states of the skyrmion crystal. We have computed the band structure in the nanoribbon geometry, where the ribbon is extended in the $x$ direction, while it has a finite width along the $y$ direction. The results are shown in Figure \ref{Edge} for three cases of $\alpha_R$. In particular, we focus on the lowest two bands and the edge states that occur in the midgap region between the two bands. The Chern numbers for the two lowest bulk subbands change from $C = -1$ to 0 to +1, as indicated in the top panel. The top panel also shows the edge states which are color coded red/green, corresponding to states localized on the bottom/top edge of the ribbon as seen from the middle panel of the figure.
As the Chern numbers change from -1 to +1, the direction of propagation of the edge state is reversed,
which is indicated by the long arrows in the middle panel of the Figure.
For the case of $\alpha_R = -0.25$, Figure \ref{Edge}(b), the lowest bulk bands are not topological in nature with zero Chern numbers, and the two edge states are degenerate in energy and constitute ordinary bound states.

\begin{figure} [tbh!]
    \centering
\includegraphics[width=1.0\linewidth]{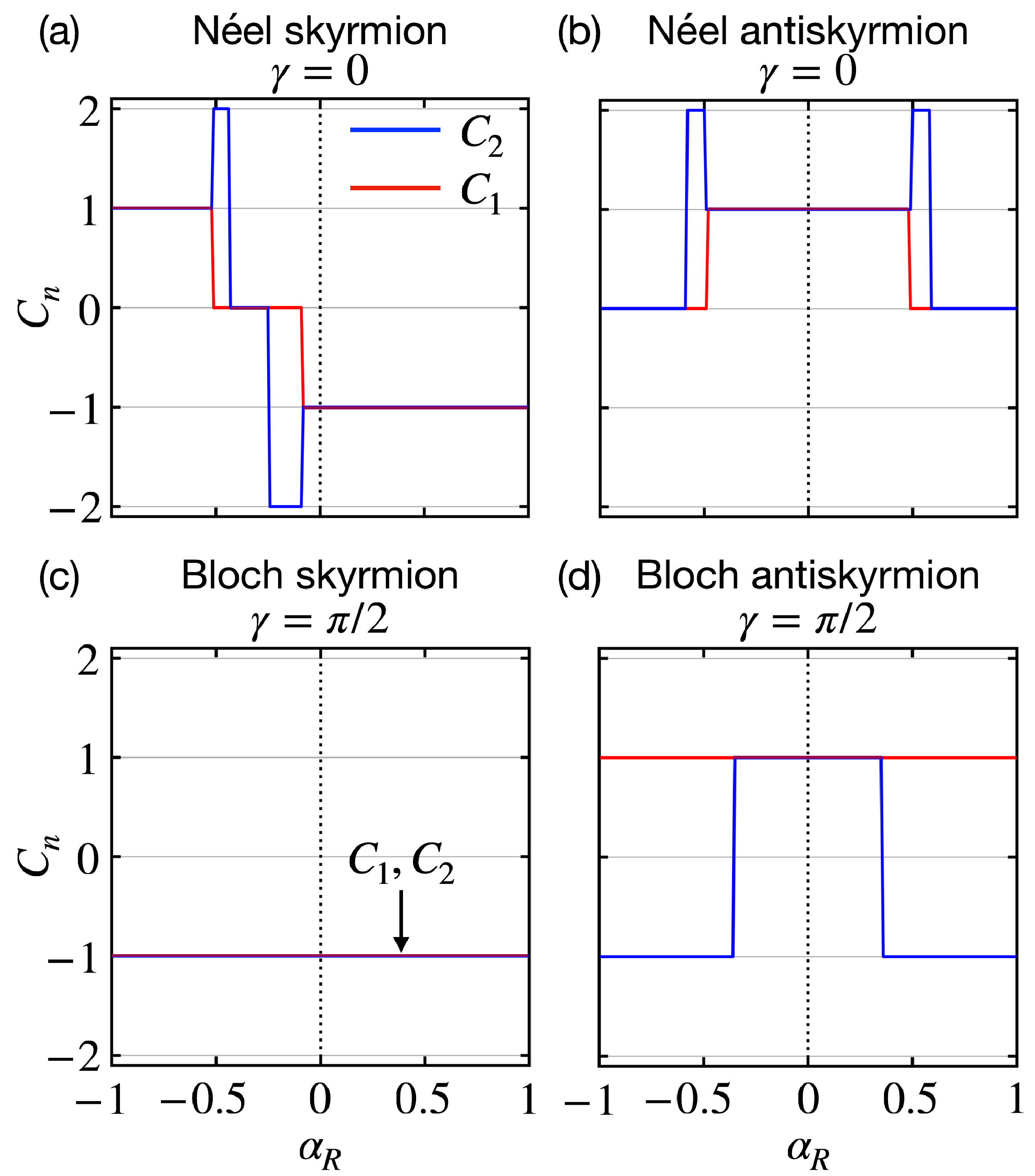}
\caption{Variation of the Chern numbers for the lowest two bands for four types of the skyrmion crystal as the Rashba SOC strength $\alpha_R$ is changed. }
    \label{Chern}
\end{figure}

{\it Band topology}.
The Chern numbers of the individual subbands change as the strength of the Rashba SOC $\alpha_R$ is varied, but always remaining an integer.
The computed Chern numbers for the lowest two bands for four skyrmion types are shown in Figure \ref{Chern}. 
The change in the Chern numbers is related to the change in the emergent magnetic field with $\alpha_R$.
Except for the N\'eel skyrmion case, the Chern numbers are symmetric with respect to the change of the sign of $\alpha_R$, which 
can be understood in terms of the emergent magnetic fields, Eqs. (\ref{sk-rb-mag}-\ref{ask-rb-mag}), and the symmetry of the square lattice as discussed below.  

(1) For the case of Bloch skyrmion (with helicity $\gamma = \pi/2$), the Chern numbers do not change with $\alpha_R$ as seen from Figure \ref{Chern} (c), since the $\cos \gamma = 0$ factor in the magnetic field expression Eq. (\ref{sk-rb-mag}) removes any magnetic field dependence of $\alpha_R$.

(2) For the antiskyrmions with any helicity, the Chern numbers are independent of the sign of $\alpha_R$ as seen from Figure \ref{Chern} (b) and (d). This is because irrespective of the helicity for the antiskyrmions, the magnetic field, Eq. (\ref{ask-rb-mag}), is a $90^\circ$ rotation of itself when the sign of $\alpha_R$ is reversed due to the $\cos(2\alpha-\gamma)$ factor, while the square lattice has the same four-fold symmetry as well. 

(3) Unlike antiskyrmions, for the case of skyrmions,  the Chern numbers for $\pm \alpha_R$ are not symmetric in general. This is because as Eq. (\ref{sk-rb-mag}) shows, the magnetic fields for $\pm \alpha_R$ are different and unrelated by any symmetry. An exception is the trivial case of $\cos \gamma = 0$, where the Rashba contribution to the magnetic field Eq. (\ref{sk-rb-mag})) is zero, which makes the Chern numbers entirely independent of $\alpha_R$, as seen from Figure \ref{Chern} (c).

(4) Of the eight skyrmion types we have considered in this paper, Figure \ref{Chern} shows the Chern numbers only for the four cases. 
The other four cases in Figure \ref{skyrmion} can be obtained from these results as well. Also, all properties for the Weyl SOC and Dresselhaus SOC including the subband Chern numbers can be expressed in terms of the results for the Rashba SOC as discussed later in the paper in Section \ref{DW}.

\begin{figure}
    \centering
    \includegraphics[width=1.0\linewidth]{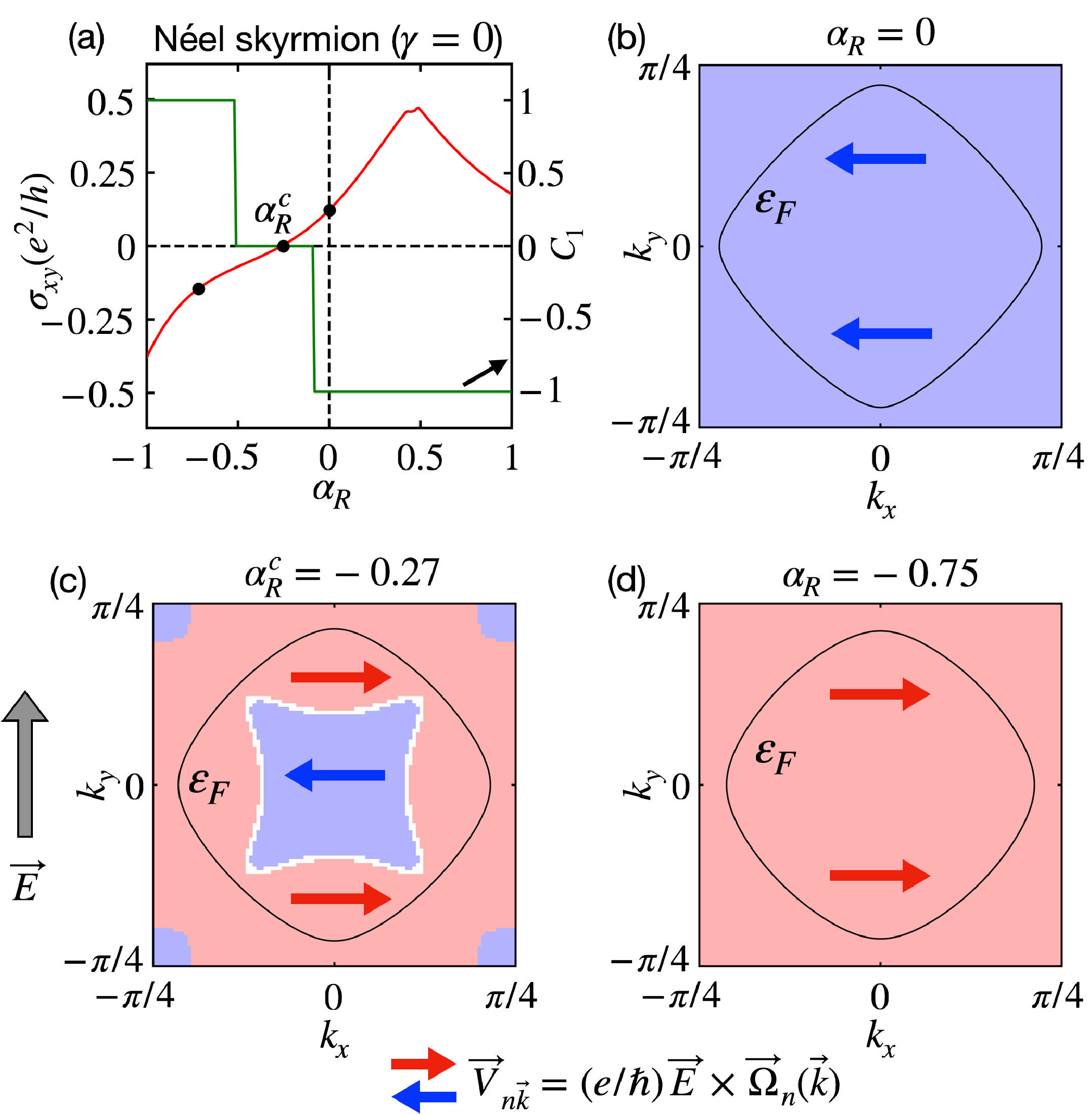}
    \caption{THC in the presence of the Rashba SOC term for the lowest subband.
    (a) The green line shows the quantized Chern number for the  {\it completely} filled lowest subband ($C_1$) as a function of $\alpha_R$, which switches between zero and $\pm 1$, leading to a quantized THC, $\sigma_{xy} = -e^2/\hbar \times C_1$. The red line shows the THC for the {\it half-filled} lowest subband indicating that it can be continuously tuned including a sign reversal at the crossover value $\alpha_R^c$. 
    (b)-(d) Berry curvatures $\Omega^z(\bold k)$ in the Brillouin zone for three different values of $\alpha_R$ 
     (corresponding to the three black dots in (a)) showing
    positive (red) and negative (blue) regions of $\Omega^z(\bold k)$. The blue/red arrows show the anomalous velocities $ e \hbar^{-1}\  \bold E \times \Omega^z(\bold k)$ for the blue/red regions (color coded). The THC, which is the sum of $\Omega^z(\bold k)$ over the occupied states,  becomes zero for  $\alpha_R^c \approx -0.27$ (Figure (c)) as the positive Berry curvature part cancels the negative part. The circles indicate the Fermi surface in each case.
    }
    \label{im3}
\end{figure}

\section{Topological Hall Effect with partially filled subbands}

The abrupt change of the Chern numbers with  $\alpha_R$ 
leads to the quantized THC for the case of completely filled subbands.
For the partially filled subbands, one gets a continuous change of the THC and the remarkable result that the sign of the THC can be flipped by a small change in $\alpha_R$.
We illustrate this here for the case where the lowest subband is half filled.

The computed $\sigma$ with the lowest subband half filled is shown in Figure~\ref{im3}(a) as a function of $\alpha_R$ for the N\'eel skyrmion, which shows a change of sign as $\alpha_R$ varies through the crossover point $\alpha_R \approx -0.27$. This leads to the remarkable result that an applied electric field can reverse the polarity of the Hall current by tuning $\alpha_R$. 

The results can be understood in terms of the change in the Berry curvatures as $\alpha_R$ is varied.
When $\alpha_R$ is varied from the value -0.75 to zero, the Chern number of the lowest subband changes from $C_1 = 1$ to $C_1 = -1$, which results in the Berry curvature $\Omega_1^z (\boldsymbol{k})$ changing from a positive value to a negative value over the entire Brillouin zone as seen from Figure \ref{im3} (b) and (d).
For intermediate values of $\alpha_R$, the positive regions slowly change into negative regions and at the crossover point, they exactly cancel each other and the THC $\sigma$, which is the sum of the Berry curvature over the occupied states, becomes zero.

\begin{figure}
    \centering
    \includegraphics[width=0.8\linewidth]{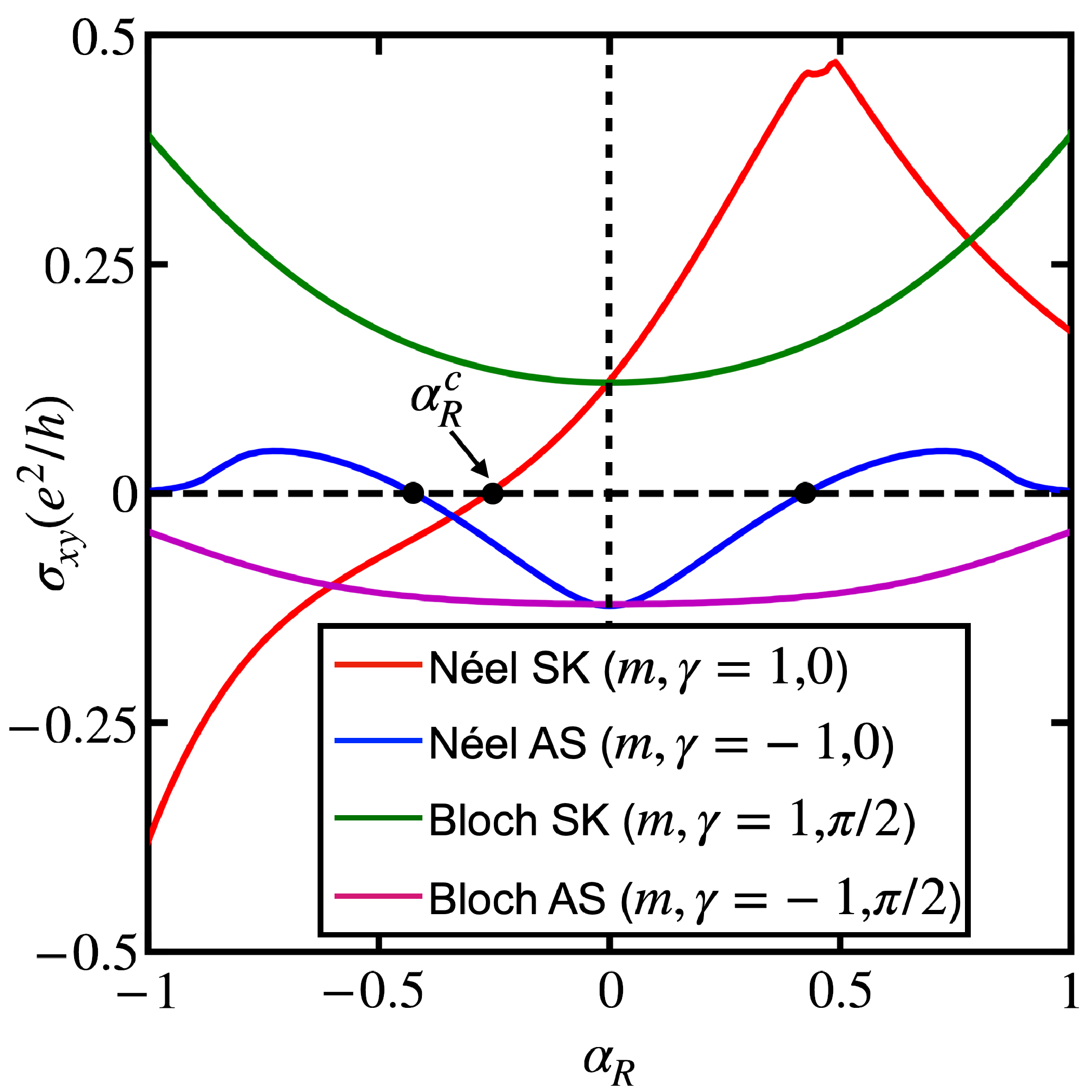}
    \caption{Variation of $\sigma_{xy}$ with the Rashba SOC strength $\alpha_R$ for four different skyrmion types, with $\sigma_{m, \gamma}$ denoting the four cases, where $m$ is the vorticity and $\gamma$ is the helicity. Here, the lowest subband is half-filled, but similar results are obtained for any partially-filled case. For completely filled subbands, $\sigma_{xy}$ changes abruptly as shown in Figure  \ref{Chern}.
    For the N\'eel skyrmion and antiskyrmion, there is a sign reversal in $\sigma_{xy}$, where changing $\alpha_R$ through the crossover value $\alpha_R^c$ changes its sign, so that the Hall current reverses in sign as illustrated in Figure \ref{im1}. The crossover points are indicated by black dots.
}
    \label{im5}
\end{figure}

We have studied the THC for the other skyrmion types as well. The results are shown in Figure \ref{im5}. A similar sign change of $\sigma$ is seen for the N\'eel antiskyrmion. For the case of the Bloch skyrmion, the Rashba SOC term does not affect the magnetic field (see Eq. (\ref{sk-rb-mag}) and as a result the Chern number for the lowest subbands remain always -1. There is no change of sign for $\sigma$ in this case. For the case of the Bloch antiskyrmion, there is no polarity change within the range of $\alpha_R$ values shown in Figure \ref{im5}, but may happen when Figure \ref{im5} is larger.

\begin{table*}
\caption{THC in the presence of Dresselhaus and Weyl SOC, expressed in terms of the THC for the Rashba SOC, with $\bar{\alpha}$ being the SOC strength in each case. 
As the Table shows, the THC for all cases can be expressed in terms of just four conductivities, viz., $\sigma_{1,0} \ (\bar{\alpha})$, $\sigma_{-1,0}\  (\bar{\alpha})$, $\sigma_{1,\pi/2}\  (\bar{\alpha})$, and $\sigma_{-1,\pi/2} \ (\bar{\alpha})$, where $\sigma_{m,\gamma} \ (\bar{\alpha}) \equiv \sigma_{m,\gamma}^R \ (\bar{\alpha})$  is the THC with Rashba SOC.
For example, the table shows that $\sigma^W_{1,0} (\bar{\alpha}) = \sigma^R_{1,\pi/2} (-\bar{\alpha})$,
which is the first result listed under Weyl.
The four independent conductivities have been plotted in Figure \ref{im5} for the case when the lowest subband is half-filled. However, the correspondence is valid for any filling, whether we have completely filled or partially filled subbands.
}
\setlength{\tabcolsep}{10pt}
\renewcommand{\arraystretch}{1.5}
\begin{center}
\begin{tabular}{l l l| l l l}
\hline
\hline
\multicolumn{3}{c|}{Skyrmion type} & Rashba & Dresselhaus & Weyl\\
Vorticity & \multicolumn{2}{c|}{Helicity} 
 & $\sigma_{m,\gamma}^R(\bar{\alpha})$ & $\sigma_{m,\gamma}^D(\bar{\alpha})$ & $\sigma_{m,\gamma}^W(\bar{\alpha})$ \\
\hline
\multirow{4}{*}{\parbox{2cm}{skyrmion \\ $m=1$}} & N\'eel & $\gamma=0$ & $\sigma_{1,0}(\bar{\alpha})$ & $-\sigma_{-1,\pi/2}(\bar{\alpha})$ & $\sigma_{1,\pi/2}(-\bar{\alpha})$ \\
& & $\gamma=\pi$ & $\sigma_{1,0}(-\bar{\alpha})$ & $-\sigma_{-1,\pi/2}(-\bar{\alpha})$ & $\sigma_{1,\pi/2}(\bar{\alpha})$ \\
\cline{2-6}
& Bloch & $\gamma=\pi/2$ & $\sigma_{1,\pi/2}(\bar{\alpha})$ & $-\sigma_{-1,0}(\bar{\alpha})$ & $\sigma_{1,0}(\bar{\alpha})$ \\
& & $\gamma=-\pi/2$ & $\sigma_{1,\pi/2}(-\bar{\alpha})$ & $-\sigma_{-1,0}(-\bar{\alpha})$ & $\sigma_{1,0}(-\bar{\alpha})$ \\
\cline{1-6}
\multirow{4}{*}{\parbox{2cm}{antiskyrmion \\ $m=-1$}} & N\'eel & $\gamma=0$ & $\sigma_{-1,0}(\bar{\alpha})$ & $-\sigma_{1,\pi/2}(\bar{\alpha})$ & $\sigma_{-1,\pi/2}(-\bar{\alpha})$ \\
& & $\gamma=\pi$ & $\sigma_{-1,0}(-\bar{\alpha})$ & $-\sigma_{1,\pi/2}(-\bar{\alpha})$ & $\sigma_{-1,\pi/2}(\bar{\alpha})$ \\
\cline{2-6}
& Bloch & $\gamma=\pi/2$ & $\sigma_{-1,\pi/2}(\bar{\alpha})$ & $-\sigma_{1,0}(\bar{\alpha})$ & $\sigma_{-1,0}(\bar{\alpha})$ \\
& & $\gamma=-\pi/2$ & $\sigma_{-1,\pi/2}(-\bar{\alpha})$ & $-\sigma_{1,0}(-\bar{\alpha})$ & $\sigma_{-1,0}(-\bar{\alpha})$ \\
\hline
\hline
\end{tabular}
\end{center}
\label{thc-all}
\end{table*}

\section { Dresselhaus and Weyl SOC}
\label{DW}

Apart from the Rashba SOC, one could also have a Dresselhaus or a Weyl type of SOC. While the Rashba SOC arises at the solid surface due to the surface inversion asymmetry and can be induced by an applied electric field, the Dresselhaus SOC arises due to the bulk inversion asymmetry. However, it has been demonstrated that 
the Dresselhaus SOC can be introduced and tuned in the 2D systems via interface engineering \cite{Rashba-Dresselhaus-2016}. Furthermore, it has been recently pointed out that depending on the crystalline symmetry of the underlying material, it is possible to have Rashba, Dresselhaus, or Weyl type SOC, even at the $\Gamma$ point of the Brillouin zone\cite{Zunger}. In view of this, for the sake of completeness, we also consider the effect of the Dresselhaus and Weyl type SOC. The SOC terms for these cases are written in momentum space as
$  \mathcal{H}_{D/W}^{\rm SOC} = 
   \alpha_{D/W} (\sigma_x k_x \mp \sigma_y k_y) 
$
with the corresponding lattice version in real space
\begin{equation}
 \mathcal{H}_{D/W}^{\rm SOC} = 
     i \alpha_{D/W} \sum_{\langle ij \rangle} 
     c_{i}^{\dagger} (\sigma_x  r^x_{ij} \mp \sigma_y  r^y_{ij}) c_{j} + H. c., 
     \end{equation} 
where the $ - (+)$ sign corresponds to the Dresselhaus (Weyl) SOC.
The corresponding full Hamiltonians, similar to Eq. (\ref{local-inf-model}) for the Rashba case, are given by
\begin{eqnarray}
  \mathcal{H}_D &=& -t\sum_{\langle ij \rangle} \cos{\frac{\theta_{ij}}{2}}e^{i a_{ij}} \ d_i^\dagger d_j \nonumber \\
    &-&  i \alpha_D \sum_{\langle ij \rangle} \big[ e^{i\phi_j}(r_{ij}^x + i r_{ij}^y) \cos{\frac{\theta_i}{2}}\sin{\frac{\theta_j}{2}} \nonumber \\
    &+& e^{-i\phi_i}(r_{ij}^x - i r_{ij}^y) \cos{\frac{\theta_j}{2}}\sin{\frac{\theta_i}{2}} \big] \ d_i^\dagger d_j + H. c.,
    \label{ds-inf}
\end{eqnarray}
\begin{eqnarray}
  \mathcal{H}_W &=& -t\sum_{\langle ij \rangle} \cos{\frac{\theta_{ij}}{2}}e^{i a_{ij}} \ d_i^\dagger d_j \nonumber \\
   &-&  i \alpha_W \sum_{\langle ij \rangle} \big[ e^{i\phi_j}(r_{ij}^x - i r_{ij}^y) \cos{\frac{\theta_i}{2}}\sin{\frac{\theta_j}{2}} \nonumber \\
    &+& e^{-i\phi_i}(r_{ij}^x + i r_{ij}^y) \cos{\frac{\theta_j}{2}}\sin{\frac{\theta_i}{2}} \big] \ d_i^\dagger d_j + H. c.,
    \label{wl-inf}
\end{eqnarray}
where again $d, d^\dagger$ are the spinless fermion operators as before, since we are taking $J \rightarrow \infty$.

As for the Rashba case, the net magnetic field seen by the electron may be determined by taking an infinitesimal square loop around a spatial point and computing the flux through it from the Peierls phase accumulated by the electron while travelling around the loop. The results are:  
\begin{widetext}
\begin{eqnarray}
    B^D_{SK}(r,\alpha) &=& \frac{ \pi}{2\lambda r}\sin\frac{\pi r}{\lambda} + \frac{\alpha_D }{t} ( \frac{\pi}{\lambda}\cos{\frac{\pi r}{\lambda}} - \frac{1}{r} \sin\frac{\pi r}{\lambda} ) \sin (2\alpha + \gamma), \ \ \ \ {\rm (Dresselhaus/ skyrmion)} \\
    B^D_{AS}(r) &=& -\frac{ \pi}{2\lambda r}\sin\frac{\pi r}{\lambda} + \frac{\alpha_D }{t} ( \frac{\pi}{\lambda}\cos{\frac{\pi r}{\lambda}} + \frac{1}{r} \sin\frac{\pi r}{\lambda} ) \sin \gamma, \ \ \ \ \ \ {\rm (Dresselhaus/ antiskyrmion)} \\
    B^W_{SK}(r) &=& \frac{ \pi}{2\lambda r}\sin\frac{\pi r}{\lambda} - \frac{\alpha_W }{t} ( \frac{\pi}{\lambda}\cos{\frac{\pi r}{\lambda}} + \frac{1}{r} \sin\frac{\pi r}{\lambda} ) \sin \gamma, \hspace{25mm} {\rm (Weyl/ skyrmion)} \\
    B^W_{AS}(r, \alpha) &=& - \frac{ \pi}{2\lambda r}\sin\frac{\pi r}{\lambda} - \frac{\alpha_W }{t} ( \frac{\pi}{\lambda}\cos{\frac{\pi r}{\lambda}} - \frac{1}{r} \sin\frac{\pi r}{\lambda} )  \sin(\gamma - {2\alpha}), \ \ \ \ \ {\rm (Weyl/ antiskyrmion)} 
\end{eqnarray}
\end{widetext}
for $r < \lambda$ and zero otherwise. The magnetic fields are in units of  
 $\Phi_0/ (2\pi)$, where
$\Phi_0$ is the flux quantum.

We note that in all cases, the magnetic field expressions can be written as sum of two magnetic fields.  The first one, arising out of the skyrmion spin texture, can be expressed as the well-known triple spin product\cite{nagaosa-review}, while the second one
due to the Rashba/Dresselhaus/Weyl SOC can be expressed as spin derivatives. The corresponding magnetic fields, in units of $ \Phi_0/(2\pi)$,  are
\begin{equation}
    B ~= -\frac{1}{2} \Vec{S} \cdot (\partial_x \Vec{S} \times \partial_y \Vec{S}), ~~~~~~~~ {\rm (Sk/AS)}
    \label{B1}
\end{equation}
\begin{equation}
B_R= -\frac{\alpha_R }{t} ( \partial_x S_x + \partial_y S_y),  ~~~~~~~~ {\rm (Rashba)}
\label{B2}
\end{equation}
\begin{equation}
B_D=~ \frac{\alpha_D }{t} ( \partial_x S_y + \partial_y S_x),  ~~~ {\rm (Dresselhaus)}
\label{B3}
\end{equation}
\begin{equation}
B_W= - \frac{\alpha_W }{t} ( \partial_x S_y - \partial_y S_x).  ~~~~~~~~ {\rm (Weyl)} 
\label{B4}
\end{equation}
The magnetic field expressions for the Rashba and Dresselhaus cases, Eqs. (\ref{B2}) and (\ref{B3}), were obtained earlier\cite{Rashba-magnetic-field}, which agree with our expressions except for an overall sign for the Dresselhaus case. However, the correctness of our sign is confirmed from the fact that the relations between the THC's presented in Table I, which we have derived analytically from the mapping of the Hamiltonians and independently checked from direct numerical calculations of the THC, would not agree otherwise. Eqs. (\ref{B1} - \ref{B4}), in the polar coordinate form, are provided in section III of the Supplemental Material \cite{supplementary}.

{\it Mapping of the Hamiltonians.}
 It turns out that the Hamiltonian for the three cases of SOC (Rashba SOC, Dresselhaus SOC, or Weyl SOC) can be mapped into one another for the skyrmion types considered in the paper. As a result, the various physical quantities, in particular, the emergent magnetic fields as well as the THC are 
 related to one another.
The THC's can be expressed, as summarized in Table \ref{thc-all}, in terms of just four quantities, viz., $\sigma_{1,0} \ (\bar{\alpha})$, $\sigma_{-1,0} \ (\bar{\alpha})$, $\sigma_{1,\pi/2} \ (\bar{\alpha})$, and $\sigma_{-1,\pi/2} \ (\bar{\alpha})$, where  $\sigma_{m,\gamma} \ (\bar{\alpha}) $ is the THC in the presence of the Rashba SOC of strength $\bar{\alpha}$, and $m$ and  $\gamma$ are the vorticity and the helicity, respectively.
These relations can be understood by noting the following points:

(1) For the THC in two cases to be the same, one must check that the emergent magnetic fields are the same, which is a necessary condition. 
However, this is not sufficient, since the effective hopping magnitude between two lattice sites can still be different even though the magnetic fields are the same. As an example, considering the THC $\sigma_{1, \pi/2}$ for the Rashba SOC case (green curve in Figure \ref{im5}), the magnetic field is independent of the Rashba SOC strength $\alpha_R$ due to the $\cos \gamma = 0$ factor (see Eq. (\ref{sk-rb-mag})). Nevertheless, the THC does change with $\alpha_R$ because the hopping amplitudes between two sites are functions of both $t$ and $\alpha_R$.

(2) The Hamiltonians possess certain symmetries. It is easy to see that in all three cases $\mathcal{H}_R$,  $\mathcal{H}_D$, and $\mathcal{H}_W$ (Eqs. (\ref{local-inf-model}), (\ref{ds-inf}), and (\ref{wl-inf})), when the sign of the SOC strength $\alpha$ is flipped and the helicity $\gamma$ is changed by $\pi$ simultaneously, the Hamiltonian remains unchanged. This leads to identical band structures and all other electronic properties including transport, so that we have the equality:
$\sigma_{m, \gamma}^X (\bar{\alpha})= \sigma_{m, \gamma \pm \pi}^X (-\bar{\alpha})$, where $X = R, D$, or $ W$.
 
\begin{figure}
    \centering
    \includegraphics[width=1.0\linewidth]{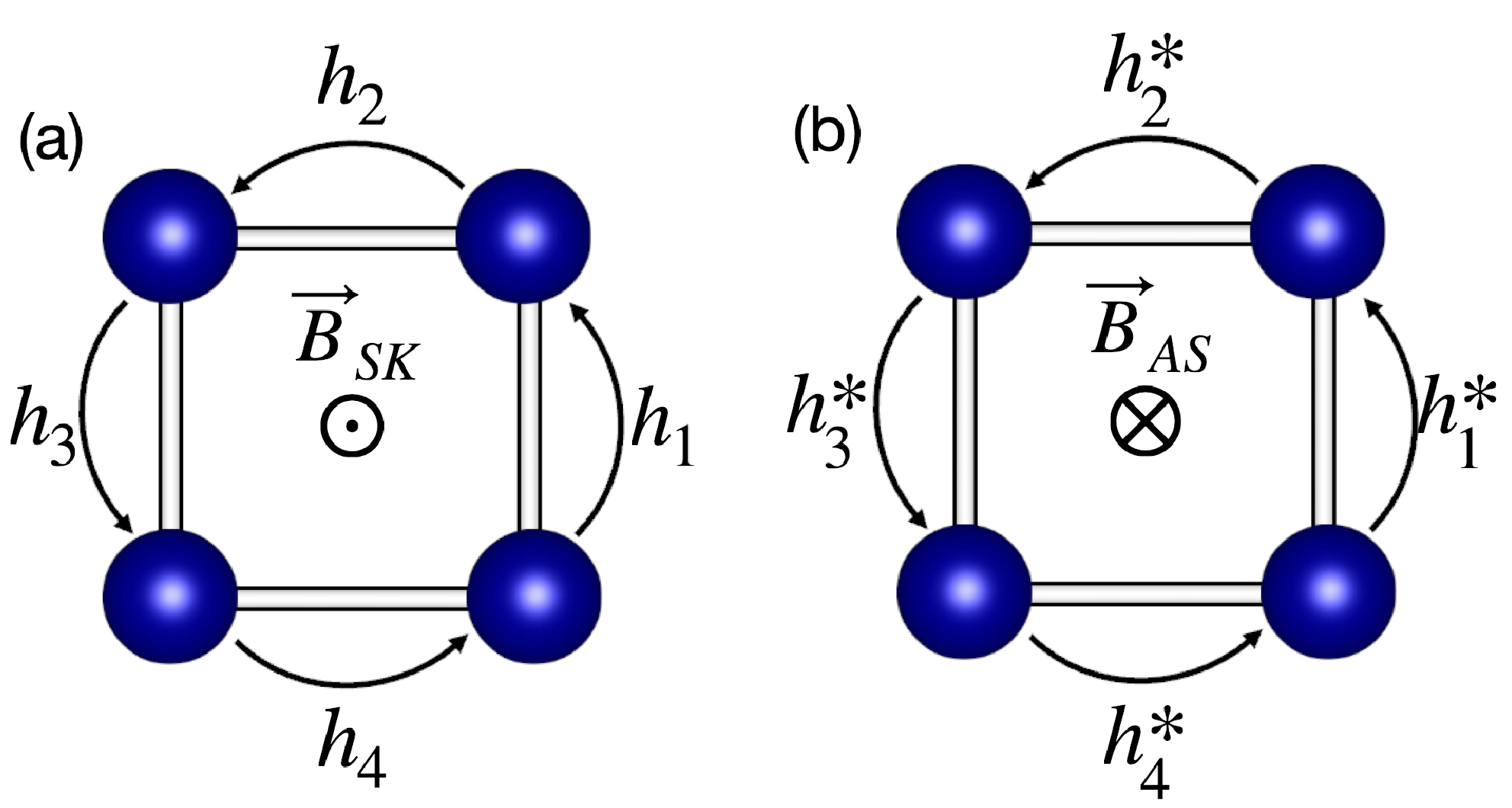}
    \caption{
    Equivalence of the Hamiltonians in the skyrmion crystal between the two cases: 
    (a) N\'eel skyrmion with $\gamma = 0$ in the presence of Rashba SOC and (b) Bloch antiskyrmion with $\gamma = \pi/2$ in the presence of Dresselhaus SOC. As discussed in the text, the effective hopping 
    $h_{ij}$ (denoted in the figure by  $h_n$), which includes both the direct hopping and the SOC term, is complex conjugate between the two cases.
    Thus, the emergent magnetic fields are the same but in the opposite directions as shown, which in turn leads to the THC, which are of the same magnitude but opposite signs between the two cases.
}
    \label{im10}
\end{figure}

(3) The Weyl Hamiltonian for a certain skyrmion type turns out to be the same as the Rashba Hamiltonian for certain other skyrmion type. For example, if we compare the Rashba Hamiltonian for the N\'eel skyrmion ($m =1, \gamma= 0$) and the Weyl Hamiltonian for the Bloch skyrmion ($m =1, \gamma= \pi/2$),
it is easily shown from Eqs. (\ref{local-inf-model}) and (\ref{wl-inf})
that the two Hamiltonians are identical \cite{supplementary}. Thus, we have  
$\sigma_{1, \pi/2}^W (\bar{\alpha})= \sigma_{1,0}^R (\bar{\alpha})$. There are several other such equalities, which have been used to construct
Table \ref{thc-all}. These relations have been explicitly verified by computing the THC's for the various cases as well. 

(4) Regarding the Rashba and Dresselhaus Hamiltonians, they  turn out to be  related 
via a complex conjugate hopping.
For instance, considering the two cases: (i) Rashba Hamiltonian for the N\'eel skyrmion ($m =1, \gamma= 0$) and (ii) Dresselhaus Hamiltonian for the Bloch antiskyrmion ($m =-1, \gamma= \pi/2$),
it is easily shown from Eqs. (\ref{local-inf-model}) and (\ref{ds-inf})
that all terms of the two Hamiltonians are the same except that the coefficients 
are complex conjugate of each other as illustrated in Figure \ref{im10}
($h_{ij} d_j^\dagger d_i$ in one case vs. $h_{ij}^* d_j^\dagger d_i$ for the other case; mathematical expressions are given in Supplementary Materials, Section IV \cite{supplementary}). This means that the effective magnetic fields that the electron sees as it hops from one site to another are simply of the opposite signs.
This in turn implies that the two THCs are negative of each other, because if two magnetic fields are applied to the same system in opposite directions, the sum of the two Hall conductivities must add up to zero. This leads to the equality $\sigma_{-1, \pi/2}^D (\bar{\alpha})= -\sigma_{1,0}^R (\bar{\alpha})$. Similarly, several other correspondences between the Rashba and Dresselhaus conductivities are found, which are used to produce Table \ref{thc-all}.

\section{Summary}

In summary, we presented a comprehensive analysis of the topological Hall effect in the skyrmion crystal in the presence of spin-orbit coupling, of Rashba, Dresselhaus, or Weyl type. We considered the square lattice as a prototype and the magnetic unit cell contained a skyrmion or an antiskyrmion of either Bloch or  N\'eel type.
Detail results for the band structure and the THC were presented for the Rashba SOC, in terms of which the other two cases can be described.

The presence of the SOC term affects the electron motion in two ways. First, it modifies the emergent magnetic field seen by the electron as it moves from site to site, and second, the hopping amplitudes are also modified. While both affect the band dispersion and the THC, the first term modifies the band topology (Chern numbers) as well.

Expressions for the emergent magnetic field were obtained for the isolated skyrmion or antiskyrmion in the presence of all three  SOC terms. These results are useful in the interpretation of the electronic band structure and the THC in the skyrmion crystal. 
The competition between the emerging magnetic field due to the skyrmion texture and the SOC term leads to a tuning of the band dispersion and, therefore, the magnitude of the THC in any type of skyrmion. For example, for the Rashba SOC case and the N\'eel skyrmion crystal, we showed that for an optimal value $\alpha_R^0$  of the  SOC strength, the combined magnetic field due to the skyrmion and the SOC term is uniform enough to produce the band structure similar to that in a uniform magnetic field, leading to prominent Hall plateaus. 

We found that the band topology, e.g., the subband Chern numbers, can be changed by varying the strength of the SOC. 
The detail behavior depends on the type of the skyrmion (vorticity and helicity) as well as the type of the SOC (Rashba, Dresselhaus, or Weyl). The change in the band topology shows up in the topological nature of the edge states. An interesting effect is that the direction of the edge current can be reversed by changing the strength of the SOC.

For partially filled subbands, the change in the band topology with the SOC strength results in the tuning of the THC. Not only its magnitude can be changed, but even the sign can be reversed in some cases, so that the direction of the Hall current flips. Explicit results were presented for Rashba SOC, where the sign flip occurs for the N\'eel  skyrmion or antiskyrmion, but not for the Bloch case. 
The sign flip was explained from the momentum-space Berry curvatures, which changes gradually from everywhere positive in the Brillouin zone to everywhere negative as the Rashba SOC strength is varied, so that at the crossover point $\alpha_R^c$, the negative and positive portions of the Berry curvatures in the Brillouin zone add up to zero, and the sign of the THC flips. This also happens for the Dresselhaus SOC and Weyl SOC, though for different types of skyrmions, which can be inferred from the mapping presented in Table \ref{thc-all}. 

When the SOC is Dresselhaus or Weyl type, all results can be mapped into the case for Rashba SOC,
because the corresponding Hamiltonians are either the same or they are related by symmetry. Keeping this in mind, we have discussed the Rashba SOC case in detail.
Results for the Dresselhaus SOC and Weyl SOC cases, including the THC, can be inferred from the appropriate mapping, which are summarized in  Table \ref{thc-all}.

Experimenters are increasingly successful in fabricating skyrmions crystals, even though their stability and manipulation hinge on a delicate interplay between magnetic interactions, external fields, and temperature. It is also well established that the strength of the Rashba SOC can be tuned via an external electric field, and also in some cases, experimenters have been able to generate and tune the Dresselhaus SOC via interface engineering.
These developments on the experimental front allow for the study of the many interesting phenomena predicted here  including the tuning of the band topology and the THC, as well as the sign reversal of the Hall current and the edge current in the nanoribbons. 
This could lead to a fertile playground for skyrmion-based spintronics applications.

\section{Acknowledgements}
This work is funded by the Department of Science and Technology, India, through Grant No. CRG/2020/004330. AM thanks MoE India for the PMRF fellowship.  SS thanks SERB India for the VAJRA fellowship.

%

\balancecolsandclearpage
\section{Supplemental Materials}
\subsection{Equivalence between the spin-orbit coupling in momentum space and the lattice version}
\label{appenB}
{\it Rashba SOC}. Here, we show that the lattice version of the Rashba spin-orbit coupling (SOC) Hamiltonian, which appears in Eq. (1) of the main text, leads to the familiar form for the Rashba term in the momentum space for the free electrons,
viz.,
\begin{equation}
    \mathcal{H}_R^k =  \alpha_R^\prime (\boldsymbol{\sigma} \times \boldsymbol{k}) \cdot \hat z
    = \alpha_R^\prime (\sigma_x k_y - \sigma_y k_x).
    \label{momentum}
\end{equation}
From the main text Eq. (1), the lattice version of the Rashba Hamiltonian is
\begin{equation}
    \mathcal{H}_R^l = - \dot{\iota} \alpha_R \sum_{i,j} c_i^{\dagger} (\sigma_x r_{ij}^y - \sigma_y r_{ij}^x) c_j,
    \label{real}
\end{equation}
where again the spin indices have been suppressed on the fermion operators $c_i^\dagger/ c_j$, $\boldsymbol{\sigma}$ is as usual the Pauli matrices, and $\boldsymbol{r}_{ij} \equiv \boldsymbol{r}_j - \boldsymbol{r}_i$ is the distance vectors between the two sites.
Taking the case of the square lattice, we show that the real space form Eq. (\ref{real}) leads to the momentum space form Eq. (\ref{momentum}) for small $k$.

Going to the momentum space with the Bloch function basis, the matrix elements of Eq. (\ref{real}) become
\begin{eqnarray}
    \mathcal{H}_R^{l,\uparrow \uparrow} &=& 0, \\
    \mathcal{H}_R^{l,\downarrow \downarrow} &=& 0, \\
    \mathcal{H}_R^{l,\uparrow \downarrow} &=& \alpha_R a(e^{\dot{\iota}k_xa}-e^{-\dot{\iota}k_xa}) - \dot{\iota}\alpha_R a(e^{\dot{\iota}k_ya}-e^{-\dot{\iota}k_ya}) \nonumber \\
    &=& 2\alpha_Ra(\sin{k_ya}+\dot{\iota}\sin{k_xa}), \\
    \mathcal{H}_R^{l,\downarrow \uparrow} &=& -\alpha_R a(e^{\dot{\iota}k_xa}-e^{-\dot{\iota}k_xa}) - \dot{\iota}\alpha_R a(e^{\dot{\iota}k_ya}-e^{-\dot{\iota}k_ya}) \nonumber \\
    &=& 2\alpha_Ra(\sin{k_ya}-\dot{\iota}\sin{k_xa}), 
\end{eqnarray}
where $a$ is the lattice constant.
To get the free-electron case, we take the small momentum limit, which immediately yields the familiar form of the Rashba Hamiltonian 
\begin{equation}
    \mathcal{H}_R^k = \alpha_R^\prime \begin{pmatrix}
        0 & k_y + \dot{\iota}k_x \\
        k_y - \dot{\iota}k_x & 0
    \end{pmatrix} = \alpha_R^\prime (\sigma_x k_y - \sigma_y k_x),
\end{equation}
where $\alpha'_R = 2\alpha_R a^2$.

{\it Dresselhaus and Weyl SOC}. Proceeding in the same way, we easily see that for the Dresselhaus and Weyl SOC, the corresponding Hamiltonian forms are equivalent for small momentum $k$:
\begin{eqnarray}
    \mathcal{H}_D^k &=&  \alpha_D^\prime (\sigma_x k_x - \sigma_y k_y), \\ 
    \mathcal{H}_D &=& - \dot{\iota} \alpha_D \sum_{i,j} c_i^{\dagger} (\sigma_x r_{ij}^x - \sigma_y r_{ij}^y) c_j, \\
    \mathcal{H}_W^k &=&  \alpha_W^\prime (\sigma_x k_x + \sigma_y k_y), \\
    \mathcal{H}_W &=& - \dot{\iota} \alpha_W \sum_{i,j} c_i^{\dagger} (\sigma_x r_{ij}^x + \sigma_y r_{ij}^y) c_j.
\end{eqnarray}

\subsection{Band structures  for two cases: Uniform magnetic field 
and the skyrmion lattice (non-uniform magnetic field) with optimal Rashba SOC}
\begin{figure}[hbt!]
    \centering
    \includegraphics[width=1\linewidth]{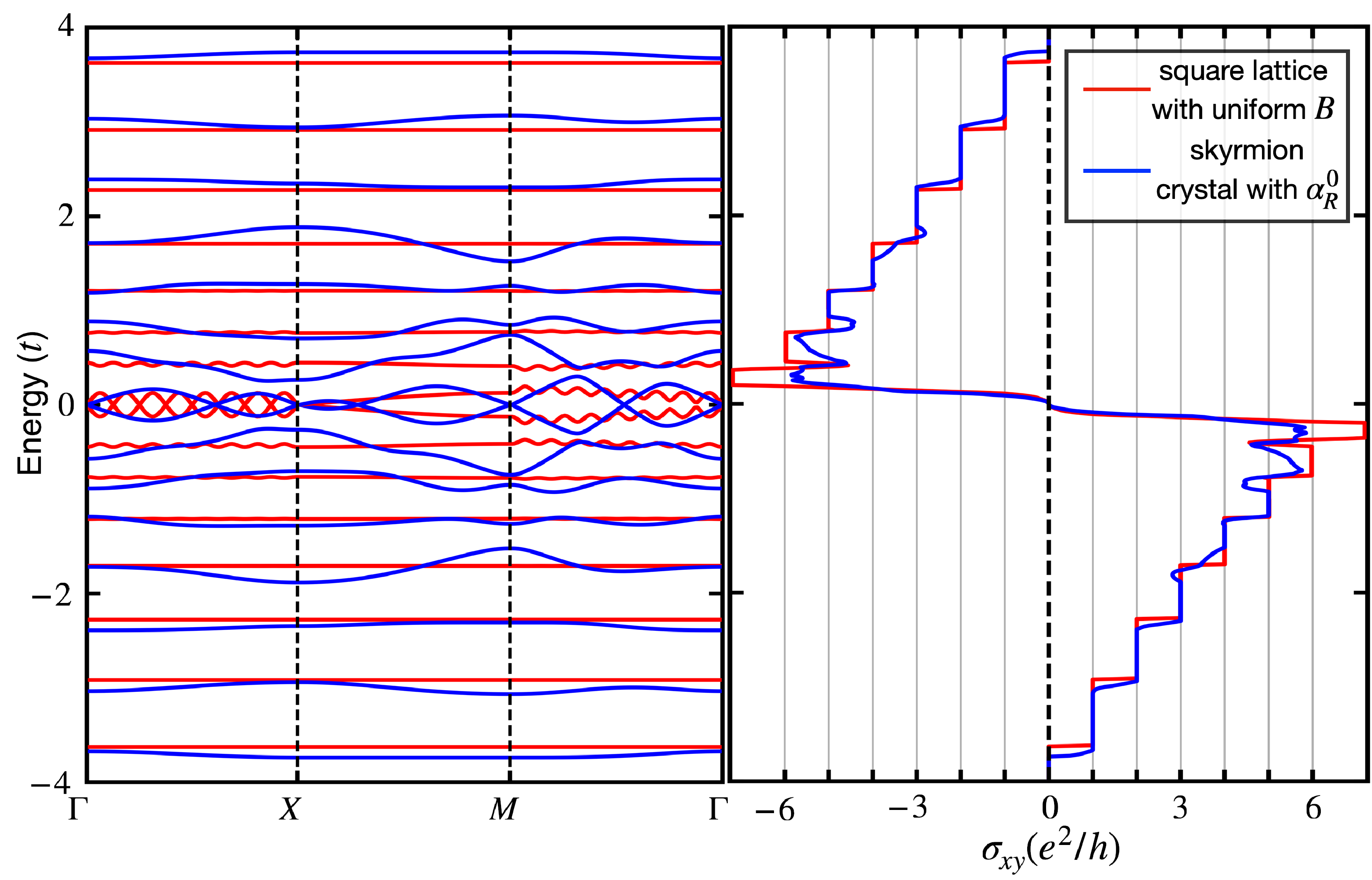}
    \caption{ Red lines show the band structure (Landau levels) and the THC $\sigma_{xy}$ for the square lattice in a uniform magnetic field.    Blue lines show the same for the  N\'eel  skyrmion crystal in the presence of Rashba SOC, with strength $\alpha_R^0$ so that the magnetic field is maximally uniform as discussed in the text. }
    \label{im1-sup}
\end{figure}
We know the SOC can modify the emergent magnetic field produced by the skyrmion spin texture. 
Note also that the Rashba SOC term, though it redistributes the magnetic field in the skyrmion, the net flux through the skyrmion remains unchanged. So, the average magnetic field in the skyrmion crystal continues to remain the same, irrespective of the strength of $\alpha_R$.
For the case of N\'eel skyrmion with Rashba SOC, we obtain one optimal value $\alpha_R^0$ where the non-uniformity in the magnetic field distribution is minimized. At $\alpha_R^0$, the average of subbands energies is compared to the energies of the Landau levels formed due to hopping of electrons in the square lattice in the presence of a uniform magnetic field, which is shown in Fig.\ref{im1-sup}(a). Not only the subband energies but also the subband dispersions are comparable to that of Landau levels in a square lattice under a uniform magnetic field, as shown in the figure. As a result, the quantization plateaus of THC become more prominent, similar to those observed in a square lattice subjected to a uniform magnetic field (the integer quantum Hall effect).

\subsection{Calculation of the emergent magnetic field in the presence of a spin texture and Rashba/Dresselhaus/Weyl SOC}

The  Rashba and other SOC terms introduce two changes in the electron motion: One, they modify the emergent magnetic field due to the skyrmion and Two, they change the magnitude of the effective hopping between the lattice sites in the crystal.  We illustrate both effects below for the Rashba SOC case.

\begin{figure}
    \centering
    \includegraphics[width=0.6\linewidth]{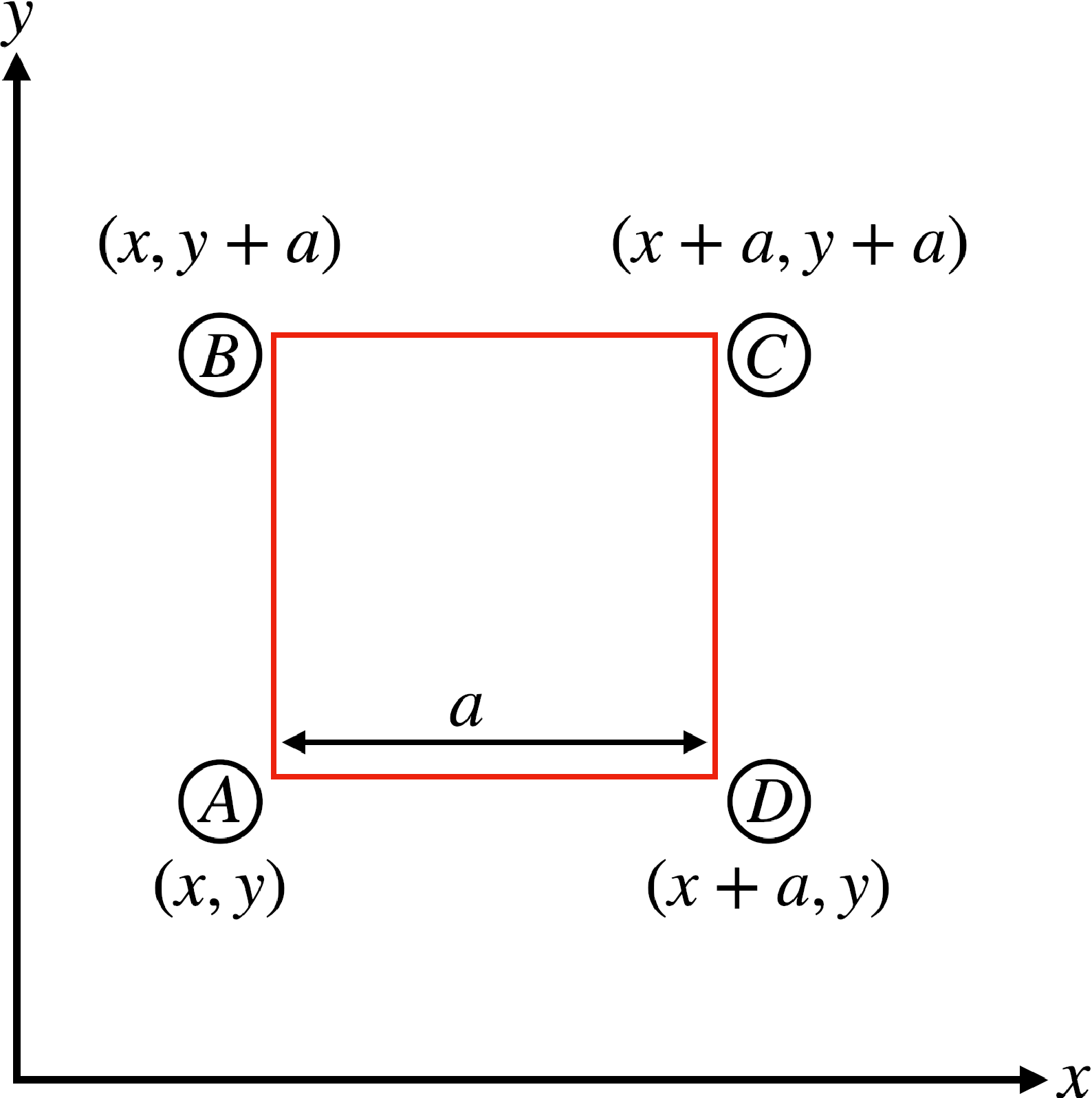}
    \caption{A square loop of size $a (a \rightarrow 0)$ used to compute the magnetic field from the complex electron hopping as the electron moves around the square in a closed loop.}
    \label{im2-sup}
\end{figure}

To calculate the emergent magnetic field, we take a square loop of side $a$ positioned at the point $(x, y)$ (see Figure \ref{im2-sup}), compute the Peierls phase as we traverse around the loop, which gives us the flux through the square, and we then take the limit $a \rightarrow 0$ to get the magnetic field. The final result for the case of the skyrmion spin texture is:
\begin{eqnarray}
    B_{SK}^R (r) &=&  \frac{\Phi_0}{2\pi} \times  \big[\frac{ \pi}{2\lambda r}\sin\frac{\pi r}{\lambda} \nonumber \\
    &-& \frac{\alpha_R }{t} ( \frac{\pi}{\lambda}\cos{\frac{\pi r}{\lambda}} + \frac{1}{r} \sin\frac{\pi r}{\lambda} ) \cos \gamma \big],
\end{eqnarray}
    which is the main result of this Section. The details of the calculation of the magnetic field are given below.

The Rashba SOC Hamiltonian, Eq. (3) of the main text, has the form 
$ \mathcal{H}_R = \sum_{\langle ij \rangle} ( h_{ij} \ d_i^\dagger d_j + h^{\ast}_{ij} \ d_j^\dagger d_i)$, where the effective hopping from site $j$ to site $i$ is written in terms of the orientation of the skyrmion spins $(\theta_i, \phi_i)$ as
\begin{eqnarray}
    h_{ij} &=& (-t \cos{\frac{\theta_i}{2}} \cos{\frac{\theta_j}{2}} - t \cos{(\phi_j - \phi_i)}\sin{\frac{\theta_i}{2}} \sin{\frac{\theta_j}{2}} \nonumber \\
    &+& \alpha_R(r_y \sin{\phi_j} + r_x \cos{\phi_j})\cos{\frac{\theta_i}{2}} \sin{\frac{\theta_j}{2}} \nonumber \\
    &-& \alpha_R(r_y \sin{\phi_i} + r_x \cos{\phi_i})\sin{\frac{\theta_i}{2}} \cos{\frac{\theta_j}{2}}) \nonumber \\
    &+&i (- t \sin{(\phi_j - \phi_i)}\sin{\frac{\theta_i}{2}} \sin{\frac{\theta_j}{2}} \nonumber \\
    &-& \alpha_R(r_y \cos{\phi_j} - r_x \sin{\phi_j})\cos{\frac{\theta_i}{2}} \sin{\frac{\theta_j}{2}} \nonumber \\
    &-& \alpha_R(r_y \cos{\phi_i} - r_x \sin{\phi_i})\sin{\frac{\theta_i}{2}} \cos{\frac{\theta_j}{2}}) \nonumber \\
    &=& t'_{ij} e^{i\delta_{ij}},
    \label{full-exp}
\end{eqnarray}
where $\delta_{ij} = \tan^{-1}(\text{Im}(h_{ij})/\text{Re}(h_{ij}))$, and in this equation we have defined $\boldsymbol{r} \equiv \boldsymbol{r}_{ij}$ for simplicity. As described in the main text, the magnetic field due to complex hopping of electron can be computed by adding the phase factors of the hopping terms around a closed loop. Here, we take a small square loop (see Figure \ref{im2-sup}) and add the four phase factors to get the magnetic flux associated with that loop and from there we obtain the magnetic field. 

We consider an infinitesimal square of side `$a$'  connecting four sites of a square at the position $(r, \alpha)$ ($(x, y)$ in the cartesian coordinaes as shown in Figure \ref{im2-sup}). In the continuum model, the   skyrmion spin orientations at the four sites are given by the spherical polar angles $(\theta, \phi)$: 
\begin{eqnarray}
    \theta_A &=& \pi\Big(1-\frac{r}{\lambda}\Big), \nonumber \\
    \phi_A &=& m \tan^{-1}\Big(\frac{y}{x}\Big) + \gamma, \\
    \theta_B &=& \pi\Big(1-\frac{r}{\lambda}-\frac{ay}{r\lambda}\Big), \nonumber \\
    \phi_B &=& m \tan^{-1}\Big(\frac{y+a}{x}\Big) + \gamma, \\
    \theta_C &=& \pi\Big(1-\frac{r}{\lambda}-\frac{a(x+y)}{r\lambda}\Big), \nonumber \\
    \phi_C &=& m \tan^{-1}\Big(\frac{y+a}{x+a}\Big) + \gamma, \\
    \theta_D &=& \pi\Big(1-\frac{r}{\lambda}-\frac{ax}{r\lambda}\Big), \nonumber \\
    \phi_D &=& m \tan^{-1}\Big(\frac{y} {x+a}\Big) + \gamma
\end{eqnarray}
We then calculate the phase factors of the complex hoppings between two consecutive sites along the closed path $A-B-C-D-A$ from Eq. (\ref{full-exp}). These four phase factors for a single-turned skyrmion $(m = 1$) can be written as
\begin{widetext}
\begin{eqnarray}
    \delta_{AB} &=& \tan^{-1}\Big[ \Big(\frac{ax}{2r^2}\Big)\Big(1-\cos{\frac{\theta_A + \theta_B}{2}}\Big) + \Big(\frac{\alpha_Ra}{t}\Big)\Big(\Big(\frac{x}{r}\sin{\frac{\theta_A + \theta_B}{2}} - \frac{axy}{r^3}\cos{\frac{\theta_A}{2}\sin{\frac{\theta_B}{2}}}\Big)\cos{\gamma} \nonumber \\
    && -\Big( \frac{y}{r}\sin{\frac{\theta_A + \theta_B}{2}} + \Big(\frac{a}{r}-\frac{ay^2}{r^3}\Big)\cos{\frac{\theta_A}{2}\sin{\frac{\theta_B}{2}}} \Big)\sin{\gamma} \Big],\\
    \delta_{BC} &=& \tan^{-1}\Big[ \Big(-\frac{ay}{2r^2}\Big)\Big(1-\cos{\frac{\theta_B + \theta_C}{2}}\Big) + \Big(\frac{\alpha_Ra}{t}\Big)\Big(-\Big(\frac{1}{r}\Big(y+a-\frac{ay^2}{r^2} \Big)\sin{\frac{\theta_B + \theta_C}{2}} - \frac{axy}{r^3}\cos{\frac{\theta_B}{2}\sin{\frac{\theta_C}{2}}}\Big)\cos{\gamma} \nonumber \\
    && -\Big( \Big(\frac{x}{r}-\frac{axy}{r^2}\Big)\sin{\frac{\theta_B + \theta_C}{2}} + \Big(\frac{a}{r}-\frac{ax^2}{r^3}\Big)\cos{\frac{\theta_B}{2}\sin{\frac{\theta_C}{2}}} \Big)\sin{\gamma} \Big],\\
    \delta_{CD} &=& \tan^{-1}\Big[ \Big(-\frac{ax}{2r^2}\Big)\Big(1-\cos{\frac{\theta_C + \theta_D}{2}}\Big) + \Big(\frac{\alpha_Ra}{t}\Big)\Big(-\Big(\frac{1}{r}\Big( x+a-\frac{ax^2}{r^2}\Big)\sin{\frac{\theta_C + \theta_D}{2}} - \frac{axy}{r^3}\sin{\frac{\theta_C}{2}\cos{\frac{\theta_D}{2}}}\Big)\cos{\gamma} \nonumber \\
    && +\Big( \Big(\frac{y}{r}-\frac{axy}{r^3}\Big)\sin{\frac{\theta_C + \theta_D}{2}} + \Big(\frac{a}{r}-\frac{ay^2}{r^3}\Big)\sin{\frac{\theta_C}{2}\cos{\frac{\theta_D}{2}}} \Big)\sin{\gamma} \Big],\\
    \delta_{DA} &=& \tan^{-1}\Big[ \Big(\frac{ay}{2r^2}\Big)\Big(1-\cos{\frac{\theta_D + \theta_A}{2}}\Big) + \Big(\frac{\alpha_Ra}{t}\Big)\Big(\Big(\frac{y}{r}\sin{\frac{\theta_D + \theta_A}{2}} - \frac{axy}{r^3}\sin{\frac{\theta_D}{2}\cos{\frac{\theta_A}{2}}}\Big)\cos{\gamma} \nonumber \\
    && +\Big( \frac{x}{r}\sin{\frac{\theta_D + \theta_A}{2}} + \Big(\frac{a}{r}-\frac{ax^2}{r^3}\Big)\sin{\frac{\theta_D}{2}\cos{\frac{\theta_A}{2}}} \Big)\sin{\gamma} \Big].
\end{eqnarray}
\end{widetext}

The magnetic flux through the loop $\Phi$ is determined from the phase factor accumulated by the electron as it traverses the loop counterclockwise from the expression 
\begin{equation}
  \Phi = \int_S \bold{B} \cdot \bold {dS}  = \oint \bold{A} \cdot \bold {dr} = 
  \frac{\hbar}{e} \times (\delta_{AB}+\delta_{BC}+\delta_{CD}+\delta_{DA}),
\end{equation}
since the electron hopping acquires the Peierls phase factor $\exp\ [  (-i e/\hbar) \int_i^j \bold{A} \cdot \bold {dr}]$ as it moves from site $i$ to $j$, where again $-e <0$ is the charge of the electron.
Summing up the phases from the above equations and taking the limit $a \rightarrow 0$, after some straightforward algebra, we get the desired result 
\begin{eqnarray}
    B_{SK}^R (r) &=& lim_{a\rightarrow 0}\frac{\Phi}{a^2} = \frac{\Phi_0}{2\pi} \times  \big[\frac{ \pi}{2\lambda r}\sin\frac{\pi r}{\lambda} \nonumber \\
    &-& \frac{\alpha_R }{t} ( \frac{\pi}{\lambda}\cos{\frac{\pi r}{\lambda}} + \frac{1}{r} \sin\frac{\pi r}{\lambda} ) \cos \gamma \big],
    \label{Bsk}
\end{eqnarray}
where, again, $\Phi_0 = h/e$ is the flux quantum.
In a similar fashion, the magnetic fields for the Dresselhaus and Weyl SOC cases can be computed,  which are given in Eqs.  (15)-(18) in the main text. 

The above-mentioned magnetic field distributions are obtained for a particular type of skyrmion, whose spin texture is defined by a linear variation of polar angle (see Eq. 4 of the main text). The general expressions for magnetic field distributions in any arbitrary skyrmion spin texture can be represented as
\begin{widetext}
\begin{equation}
    B ~= \frac{\Phi_0}{4 \pi}\frac{m}{r}\frac{d\theta}{dr}\sin{\theta}, ~~~~~~~~ {\rm (Sk/AS)}
    \label{B11}
\end{equation}
\begin{equation}
B_R= \frac{\Phi_0}{2\pi} \times  \big[ \frac{\alpha_R }{t} (\frac{d \theta}{dr} \cos{\theta} + \frac{m}{r} \sin{\theta} ) \cos{(\alpha - \phi)} \big],  ~~~~~~~~ {\rm (Rashba)}
\label{B22}
\end{equation}
\begin{equation}
B_D=~ -\frac{\Phi_0}{2\pi} \times  \big[\frac{\alpha_D }{t} (\frac{d \theta}{dr} \cos{\theta} - \frac{m}{r} \sin{\theta} ) \sin{(\alpha + \phi)} \big],  ~~~ {\rm (Dresselhaus)}
\label{B33}
\end{equation}
\begin{equation}
B_W= \frac{\Phi_0}{2\pi} \times  \big[ \frac{\alpha_W }{t} (\frac{d \theta}{dr} \cos{\theta} + \frac{m}{r} \sin{\theta} ) \sin{(\phi-\alpha)} \big].  ~~~~~~~~ {\rm (Weyl)} 
\label{B44}
\end{equation}
The total magnetic field is a sum of the skyrmion magnetic field plus the field due to the SOC. For example, if Rashba SOC is present, then the net magnetic field is $B_{\text{tot}} = B + B_R$ and so on.

 {\it Magnetic field and electron transport} -- We note that electron transport properties such as THC are not just determined by the phase factor in the hopping term $ h_{ij} \equiv t^\prime_{ij} \exp (i \delta_{ij})$, which determines the magnetic field, but also by the strength of the hopping $t^\prime_{ij}$ as well. Thus even though the magnetic field may be the same for two cases, the transport properties could be different. 

 As an example, we consider the case of the Bloch skyrmion ($\gamma = \pi/2$) in the presence of the Rashba SOC. Since $\cos \gamma = 0$, the magnetic field is unaffected by the presence of the Rasba term as seen from Eq. (\ref{Bsk}). However,  the magnitude of the four hoppings, given by Eq. (\ref{full-exp}), around the square loop do depend on the Rashba SOC. To illustrate this, we compute the hopping amplitudes for the square loop sitting on the $x$-axis at the point $(x, 0)$ far away from the origin ($x >> a$). This condition simplifies the evaluation of Eq. (\ref{full-exp}).  Note also that $a$ is the lattice constant here and is not infinitesimal, which was used in the calculation of the magnetic field above. The results are:
\begin{eqnarray}
    |t'_{AB}|^2 &=& t^2 + \Big(-t\frac{a}{x}\cos^2\frac{\pi x}{2\lambda}+\alpha_R \frac{a^2}{2x}\sin{\frac{\pi x}{\lambda}}\Big)^2, \\
    |t'_{BC}|^2 &=& \Big(-t\cos{\frac{\pi a}{2\lambda}}+\alpha_R\frac{a^2}{x}\sin{\frac{\pi a}{2\lambda}}\Big)^2 + \Big(-t(\frac{a}{x+a}-\frac{a}{x})\cos^2\frac{\pi x}{2\lambda}+\alpha_Ra \ \sin{\frac{\pi x}{\lambda}}\Big)^2, \\
    |t'_{CD}|^2 &=& t^2 + \Big(t\frac{a}{x+a}\cos^2\frac{\pi(x+a)}{2\lambda} - \alpha_R \frac{a}{2}\frac{a}{x+a} \sin{\frac{\pi(x+a)}{\lambda}}\Big)^2, \\
    |t'_{DA}|^2 &=& \Big(t\cos{\frac{\pi a}{2\lambda}}\Big)^2 + \Big(\alpha_Ra \ \sin{\frac{\pi x}{\lambda}}\Big)^2.
\end{eqnarray}
The hopping amplitudes clearly depend on the strength of the Rashba SOC $\alpha_R$, even though the magnetic field in this example remains the same irrespective of  $\alpha_R$.
\end{widetext}

\subsection{Equivalence of electronic Hamiltonian with Rashba, Dresselhaus, and Weyl SOC}
As explained in the main text, the Hamiltonian (and therefore the magnetic field as well) with Dresselhaus and Weyl SOC can be mapped to the Hamiltonian with Rashba SOC. Here, we provide explicit examples for two cases.

Case I.  N\'eel skyrmion with helicity $\gamma = 0$ and Rashba SOC vs.  Bloch skyrmion with 
helicity $\gamma = \pi/2$ and Weyl SOC.

Using Eqs. (3) and (14) of the main text, the hopping terms between two sites are written as
\begin{eqnarray}
    h_R^{ij} &=&  (-t)\cos{\frac{\theta_{ij}}{2}e^{ia_{ij}}} -i\alpha_R[((r_{ij}^y \cos{\alpha_j} - r_{ij}^x \sin{\alpha_j}) \nonumber \\
    &+& i(r_{ij}^x \cos{\alpha_j} + r_{ij}^y \sin{\alpha_j}))\cos{\frac{\theta_i}{2}}\sin{\frac{\theta_j}{2}} \nonumber \\
    &+& ((r_{ij}^y\cos{\alpha_i}-r_{ij}^x\sin{\alpha_i})\nonumber \\
    &-& i(r_{ij}^x\cos{\alpha_i} + r_{ij}^y\sin{\alpha_i})\cos{\frac{\theta_j}{2}}\sin{\frac{\theta_i}{2}})],
    \label{exp1}
\end{eqnarray}
\begin{eqnarray}
    h_W^{ij} &=&  (-t)\cos{\frac{\theta_{ij}}{2}e^{ia_{ij}}} -i\alpha_W[((r_{ij}^y \cos{\alpha_j} - r_{ij}^x \sin{\alpha_j}) \nonumber \\
    &+& i(r_{ij}^x \cos{\alpha_j} + r_{ij}^y \sin{\alpha_j}))\cos{\frac{\theta_i}{2}}\sin{\frac{\theta_j}{2}} \nonumber \\
    &+& ((r_{ij}^y\cos{\alpha_i}-r_{ij}^x\sin{\alpha_i}) \nonumber \\
    &-& i(r_{ij}^x\cos{\alpha_i} + r_{ij}^y\sin{\alpha_i})\cos{\frac{\theta_j}{2}}\sin{\frac{\theta_i}{2}})].
    \label{exp2}
\end{eqnarray}
These two Hamiltonians are identical term by term: $h_R^{ij} = h_W^{ij}$ for the same SOC strength
$\alpha_R = \alpha_W$. This means that all electronic properties are the same for the two cases including the emerging magnetic field and the THC.

Case II.  N\'eel skyrmion with helicity $\gamma = 0$ and Rashba SOC vs.  Bloch antiskyrmion with 
helicity $\gamma = \pi/2$ and Dresselhaus SOC.

The hopping term between two sites for the first scenario is already given in Eq. (\ref{exp1}). The hopping term for the second scenario from Eq. (13) of the main text is
\begin{eqnarray}
    h_D^{ij} &=&  (-t)\cos{\frac{\theta_{ij}}{2}e^{-ia_{ij}}} -i\alpha_D[((-r_{ij}^y \cos{\alpha_j} + r_{ij}^x \sin{\alpha_j}) \nonumber \\
    &+& i(r_{ij}^x \cos{\alpha_j} + r_{ij}^y \sin{\alpha_j}))\cos{\frac{\theta_i}{2}}\sin{\frac{\theta_j}{2}} \nonumber \\
    &+& ((-r_{ij}^y\cos{\alpha_i}+r_{ij}^x\sin{\alpha_i}) \nonumber \\
    &-& i(r_{ij}^x\cos{\alpha_i} + r_{ij}^y\sin{\alpha_i})\cos{\frac{\theta_j}{2}}\sin{\frac{\theta_i}{2}})].
    \label{exp3}
\end{eqnarray}
From Eq. (\ref{exp1}) and (\ref{exp3}), one can easily see that $h_R^{ij} = h_D^{ij\ast}$, which are complex conjugates of each other. So, the electronic properties are related to each other via a symmetry transformation. For instance, the emergent magnetic field is same in magnitude but opposite in direction, which makes the THC also the same in magnitude but opposite in sign.

\subsection{Quantization of topological Hall effect for a larger skyrmion}

\begin{figure}[hbt!]
    \centering
    \includegraphics[width=1\linewidth]{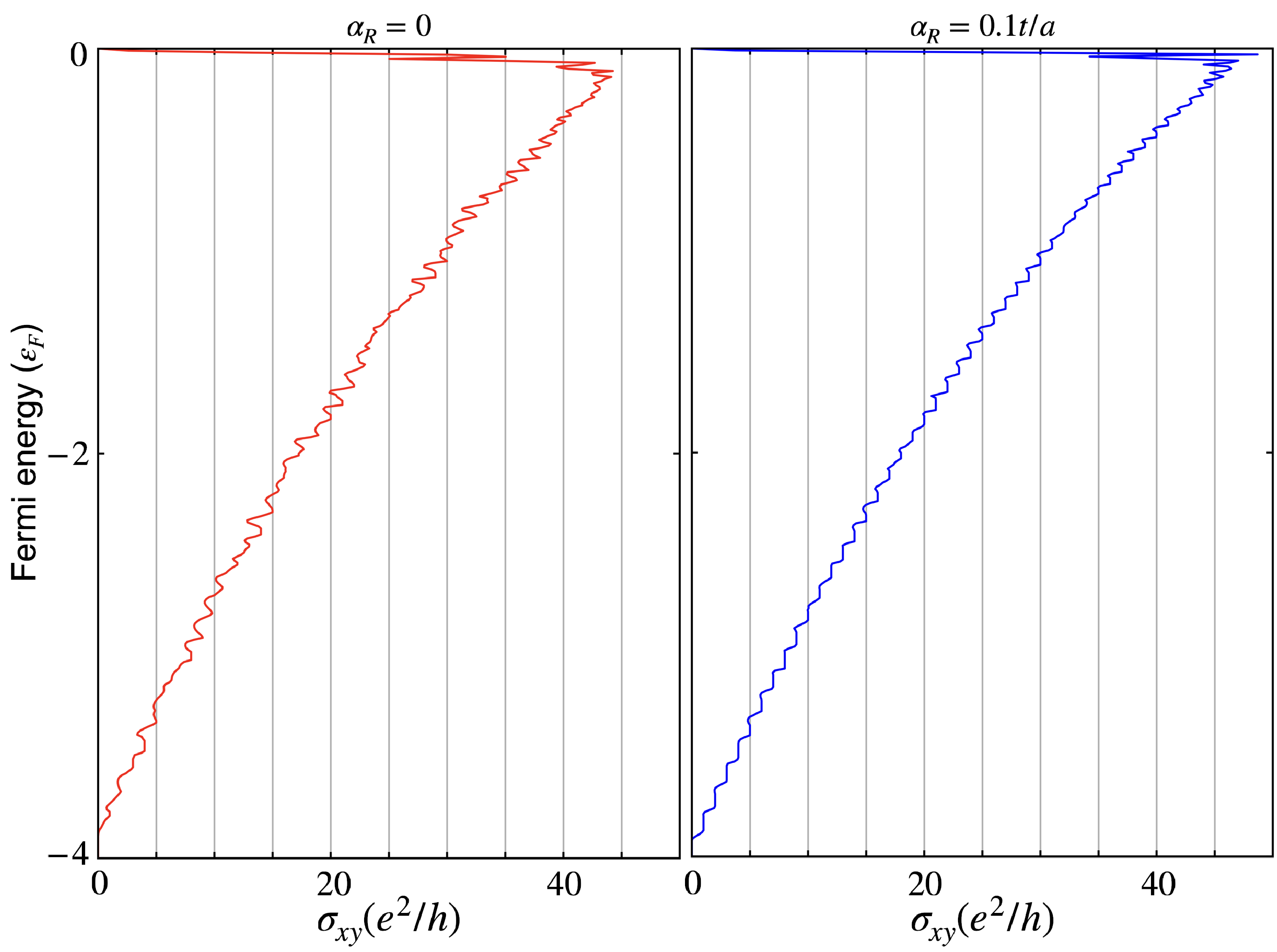}
    \caption{Quantization of THC in a N\'eel type skyrmion of size $10 \times 10$. The left column represents the THC in the absence of Rashba SOC, while the right column represents the THC with the optimal Rashba SOC. In this figure, we show the lower half of the band spectrum to maintain the clarity of the quantization plateaus. The upper half will be just opposite of the lower one, as shown in Fig. 4 of the main text.}
    \label{im3-sup}
\end{figure}

In Figure 4 of the main text, the THC is shown for a N\'eel type skyrmion of size $4 \times 4$, and the effect of Rashba SOC on the quantization plateaus of THC is presented. Since the total magnetic flux associated with a skyrmion is independent of its radius, all the properties demonstrated in the main text with a skyrmion of size $4 \times 4$ are equally valid for any size of the skyrmion. Figure \ref{im3-sup} shows the THC for a skyrmion with the magnetic unit cell size $10 \times 10$. One can see the conductivity plateaus and the same Chern numbers. The optimal strength of the Rashba SOC that leads to a well-defined Quantum Hall like plateaus can also be estimated from the expression, Eq. (7), of the main text,
to be 
$\alpha_R^0 \approx 0.1 \  t/a$. The THC with this optimal Rashba SOC is shown on the right hand side of Figure \ref{im3-sup}. From the figure, it is evident that even for larger skyrmions, a critical $\alpha_R^0$ minimizes the non-uniformity of the magnetic field distribution, and thereby enhancing the quality of the quantization plateaus.

\subsection{Topological Hall conductivity under finite Hund's coupling}
\begin{figure*}[hbt!]
    \centering
    \includegraphics[width=1\linewidth]{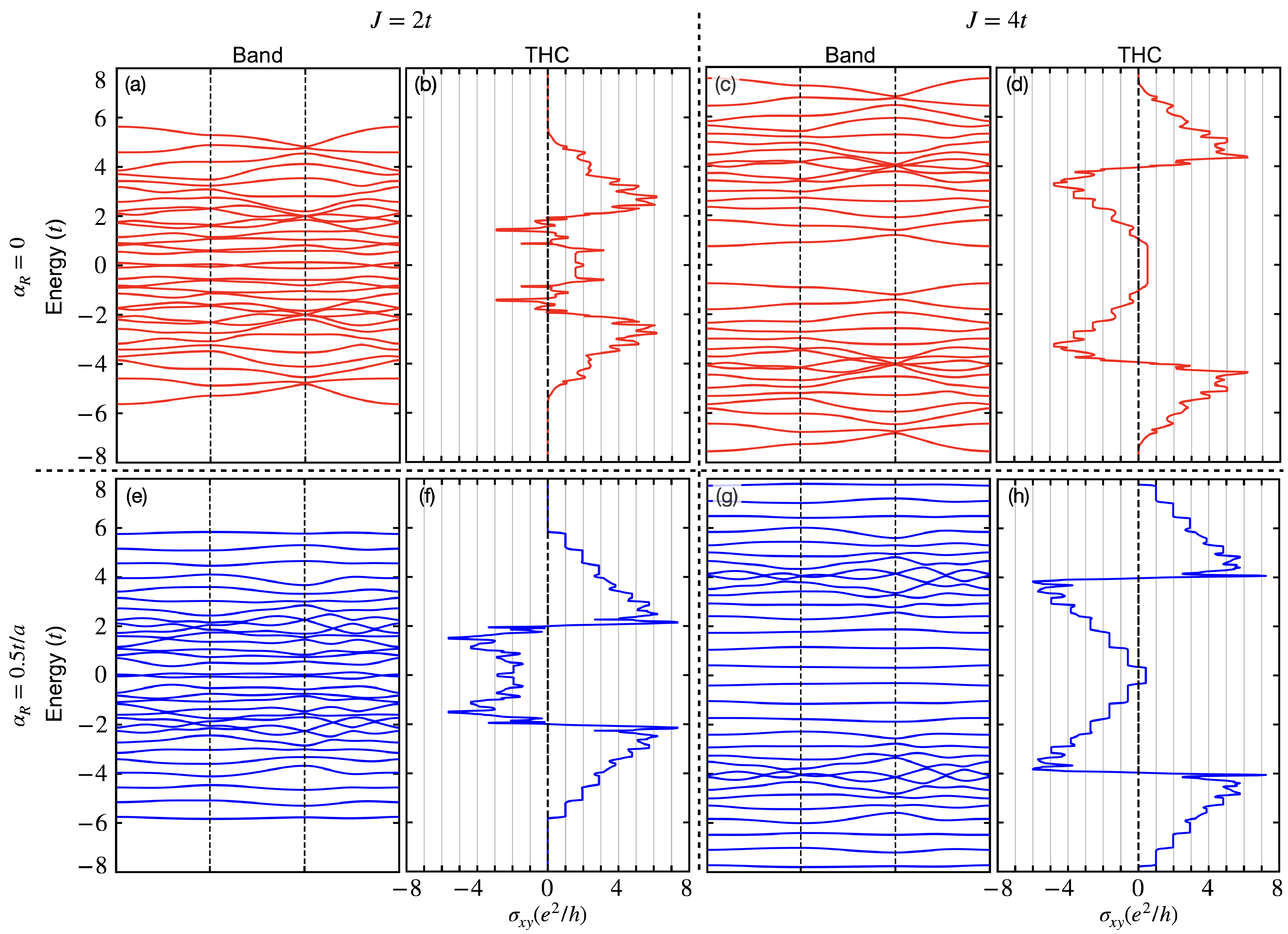}
    \caption{Topological Hall conductivity in N\'eel type square skyrmion crystal under finite Hund's coupling ($J$). The top panel represents the band structure and THC without Rashba SOC, while the bottom panel represents the same for optimal Rashba SOC (i.e., $\alpha_R = 0.5 t/a$). The left two columns show the THC for $J=2t$, where the bands of spin up and spin down block partially overlap. The quantum Hall-like plateaus are observed for the energy domain where the spin-up and spin-down bands do not overlap. In the right two columns, the THC is shown for $J=4t$, where the bands for spin up and spin down blocks are well-separated, nearly resembling the case of the infinite $J$ limit. In this case, the optimal $\alpha_R$ can produce the quantum Hall-like plateaus in any energy domain.}
    \label{im4-sup}
\end{figure*}
Since most of the materials have very high Hund's coupling ($J$) compared to the hopping strength, we consider the infinite $J$ limit for the analyses of results in the main text. In this limit, the spin-down states of electrons are ignored as their energy is $2J$ higher compared to the corresponding spin-up state, and the spinfull fermion problem converts into the spinless fermion problem. However, no additional complications will appear if the finite $J$ and thereby both the spin states are considered. Here, we show the THC of a N\'eel type skyrmion and its quantization plateaus in the presence of optimal Rashba SOC for finite Hund's coupling strength.

The Hamiltonian for itinerant electrons moving through a square skyrmion crystal in the presence of Rashba SOC under the global spin basis is written as (see Eq. 1 of the main text),
\begin{widetext}
\begin{equation}
\mathcal{H} = - t \sum_{\langle ij \rangle}  (c_{i}^{\dagger} c_{j} + c_{j}^{\dagger} c_{i})-J\sum_{i} \boldsymbol{S}_i \cdot c_{i}^{\dagger}\boldsymbol{\sigma}c_{i} - i \alpha_R \sum_{\langle ij \rangle} c_{i}^{\dagger} (\boldsymbol{\sigma} \times \boldsymbol{r}_{ij})_z c_{j} + H. c..
\label{eqA1}
\end{equation}
The spinor of an electron is defined as $|\chi\rangle = \cos(\frac{\theta}{2})|\uparrow\rangle + e^{\dot{\iota}\phi} \sin(\frac{\theta}{2})|\downarrow\rangle$. Therefore, the local spin vector can be defined as
\begin{eqnarray}
    \boldsymbol{S} &=& \langle \chi |\sigma_x|\chi \rangle \hat{x} + \langle \chi |\sigma_y|\chi \rangle \hat{y} + \langle \chi |\sigma_z|\chi \rangle \hat{z} = \sin{\theta}\cos{\phi} \hat{x} + \sin{\theta}\sin{\phi} \hat{y} + \cos{\theta} \hat{z},
\end{eqnarray}
where $\sigma_x$, $\sigma_y$, and $\sigma_z$ are Pauli matrices. To rewrite the Hamiltonian of Eq.~\ref{eqA1} in local spin basis ($\chi_{\uparrow}/\chi_{\downarrow}$), we need to perform a unitary basis transformation. In the second term of Eq.~\ref{eqA1}, the operator $\boldsymbol{S}\cdot \boldsymbol{\sigma}$ is not diagonal in $|\uparrow \rangle$ and $|\downarrow \rangle$ basis but the same is diagonal in $|\chi_{\uparrow}\rangle=\cos{(\theta/2)} |\uparrow \rangle + e^{\dot{\iota}\phi}\sin{(\theta/2)} |\downarrow \rangle$ and $|\chi_{\downarrow} \rangle=\cos{((\pi-\theta)/2)} |\uparrow \rangle + e^{\dot{\iota}(\phi+\pi)}\sin{((\pi-\theta)/2)} |\downarrow \rangle$ basis with eigenvalues $+1$ and $-1$ that means in local spin basis the Hund's coupling takes the form of $\sigma_z$. Now the unitary matrix $U$ that does the transformation $U^{\dagger}(\boldsymbol{S}\cdot \boldsymbol{\sigma})U=\sigma_z$ can be expressed as
\begin{equation}
    U = \begin{pmatrix}
    \cos{(\theta/2)} & e^{-\dot{\iota}\phi}\sin{(\theta/2)}\\
    e^{\dot{\iota}\phi}\sin{(\theta/2)} & -\cos{(\theta/2)}
    \end{pmatrix},
\end{equation}
Therefore, the creation and annihilation operator ($c_i^{\prime \dagger}$ and $c^{\prime}_i$) in local spin basis can be transformed as
\begin{equation}
    c_i = U_i c^{\prime}_i.
\end{equation}
The Hamiltonian of Eq.~\ref{eqA1} can be written in the local spin basis as
\begin{eqnarray}
    \mathcal{H} &=& - t \sum_{\langle ij \rangle}  (c^{\prime \dagger}_{i} U_i^{\dagger} U_j c^{\prime}_{j} + c_{j}^{\prime \dagger} U_j^{\dagger} U_i c^{\prime}_{i})-J\sum_{i} \boldsymbol{S}_i \cdot c_{i}^{\prime \dagger} U_i^{\dagger}\boldsymbol{\sigma}U_i c^{\prime}_{i} \nonumber - i \alpha_R \sum_{\langle ij \rangle} c_{i}^{\prime \dagger} U_i^{\dagger} (\boldsymbol{\sigma} \times \boldsymbol{r}_{ij})_z U_j c^{\prime}_{j} + H. c.. \nonumber \\
    &=& -t \sum_{\langle ij \rangle,\sigma,\sigma^{\prime}} (\langle \chi_{i, \sigma}|\chi_{j, \sigma^{\prime}} \rangle c_{i,\sigma}^{\prime \dagger} c^{\prime}_{j,\sigma^{\prime}} + \langle \chi_{j, \sigma^{\prime}}|\chi_{i, \sigma} \rangle c_{j,\sigma^{\prime}}^{\prime \dagger} c^{\prime}_{i,\sigma}) -J\sum_i c_i^{\prime \dagger} \sigma_z c^{\prime}_i - i \alpha_R \sum_{\langle ij \rangle, \sigma \sigma^{\prime}} c_{i, \sigma}^{\prime \dagger} \langle \chi_{i, \sigma} | (\boldsymbol{\sigma} \times \boldsymbol{r}_{ij})_z| \chi_{j, \sigma^{\prime}} \rangle c^{\prime}_{j, \sigma^{\prime}} + H. c.. \nonumber \\
    \label{eq2-vi}
\end{eqnarray}
From Eq. \ref{eq2-vi}, one can obtain Eq. 3 of the main text by considering the infinite Hund's coupling limit, where only the lower block of the band structure of Fig. \ref{im4-sup}(c) and (g) is considered.
\end{widetext}

The numerical computations are very straightforward. The band structure and THC for finite $J$ can be computed both from Eq. \ref{eqA1} (i.e., global spin basis) and \ref{eq2-vi} (i.e, local spin basis) as shown in Fig. \ref{im4-sup}. Here, we have considered two Hund’s coupling values: (a) $J = 2t$, where some parts of the spin-up and spin-down band blocks overlap with each other, and (b) $J = 4t$, where both the band blocks get separated. From the figure, one can see that even in the finite $J$ limit, the optimal Rashba SOC exhibits the quantum Hall-like plateaus everywhere in THC except the region where different spin bands overlap.

\end{document}